\documentclass[aps,reprint,superscriptaddress,amsmath,amssymb,prx]{revtex4-2}
\usepackage{amsmath}
\usepackage{graphicx}
\usepackage{dcolumn}
\usepackage{bm}
\usepackage{float}
\usepackage[english]{babel} 
\usepackage[usenames,dvipsnames]{xcolor}
\usepackage[colorlinks=true,citecolor=Cerulean,linkcolor=RubineRed,urlcolor=Cerulean]{hyperref}
\usepackage{algorithm}
\usepackage{algorithmic}
\newcommand{\ket}[1]{|{#1}\rangle}
\newcommand{\bra}[1]{\langle{#1}|}

\newtheorem{theorem}{Theorem}
\newtheorem{definition}[theorem]{Definition}
\DeclareMathOperator{\Tr}{Tr}
\DeclareMathOperator{\Imm}{Im}
\usepackage{mathtools,booktabs}
\DeclarePairedDelimiter\ceil{\lceil}{\rceil}

\begin{document}
	 
\title{Bulk Localised Transport States in Infinite and Finite Quasicrystals via Magnetic Aperiodicity}
\author{Dean Johnstone}
\thanks{These authors contributed equally to this work.\\ \color{Cerulean} dj79@hw.ac.uk\color{Black}; \\ \color{Cerulean} m.colbrook@damtp.cam.ac.uk}
\affiliation{SUPA, Institute of Photonics and Quantum Sciences, Heriot-Watt University, Edinburgh EH14 4AS, UK}
\author{Matthew J. Colbrook}
\thanks{These authors contributed equally to this work.\\ \color{Cerulean} dj79@hw.ac.uk\color{Black}; \\ \color{Cerulean} m.colbrook@damtp.cam.ac.uk}
\affiliation{Department of Applied Mathematics and Theoretical Physics, University of Cambridge, Wilberforce Rd, Cambridge CB3 0WA, UK}
\author{Anne E. B. Nielsen}
\affiliation{Max-Planck-Institut f\"{u}r Physik komplexer Systeme, D-01187 Dresden, Germany}
\affiliation{Department of Physics and Astronomy, Aarhus University, DK-8000 Aarhus C, Denmark}
\author{Patrik \"{O}hberg}
\affiliation{SUPA, Institute of Photonics and Quantum Sciences,
	Heriot-Watt University, Edinburgh EH14 4AS, UK}
\author{Callum W. Duncan}
\email{callum.duncan@strath.ac.uk}
\affiliation{Department of Physics and SUPA, University of Strathclyde, Glasgow G4 0NG, United Kingdom}
\affiliation{Max-Planck-Institut f\"{u}r Physik komplexer Systeme, D-01187 Dresden, Germany}
\date{\today}

\begin{abstract}
Robust edge transport can occur when charged particles in crystalline lattices interact with an applied external magnetic field. This system is well described by Bloch's theorem, with the spectrum being composed of bands of bulk states and in-gap edge states. When the confining lattice geometry is altered to be quasicrystaline, i.e.\ quasiperiodic, then Bloch's theorem breaks down. However, for the quasicrystalline system, we still expect to observe the basic characteristics of bulk states and current carrying edge states. Here, we show that for quasicrystals in magnetic fields, there is also a third option; the bulk localised transport states. These states share the in-gap nature of the well-known edge states and can support transport along them, but they are fully contained within the bulk of the system, with no support along the edge. This results in transport being possible both along the edge and within distinct regions of the bulk. We consider both finite-size and infinite-size systems, using rigorous error controlled computational techniques that are not prone to finite-size effects. The bulk localised transport states are preserved for infinite-size systems, in stark contrast to the normal edge states. This allows for transport to be observed in infinite-size systems, without any perturbations, defects, or boundaries being introduced. We confirm the in-gap topological nature of the bulk localised transport states for finite and infinite-size systems by computing common topological measures; namely the Bott index and local Chern marker. The bulk localised transport states form due to a magnetic aperiodicity arising from the interplay of length scales between the magnetic field and quasiperiodic lattice. Bulk localised transport could have interesting applications similar to those of the edge states on the boundary, but that could now take advantage of the larger bulk of the lattice. The infinite size techniques introduced here, especially the calculation of topological measures, could also be widely applied to other crystalline, quasicrystalline, and disordered models.
\end{abstract}

\maketitle

\section{Introduction}

\begin{figure*}[ht!]
	\centering
	\includegraphics[width=0.8\linewidth]{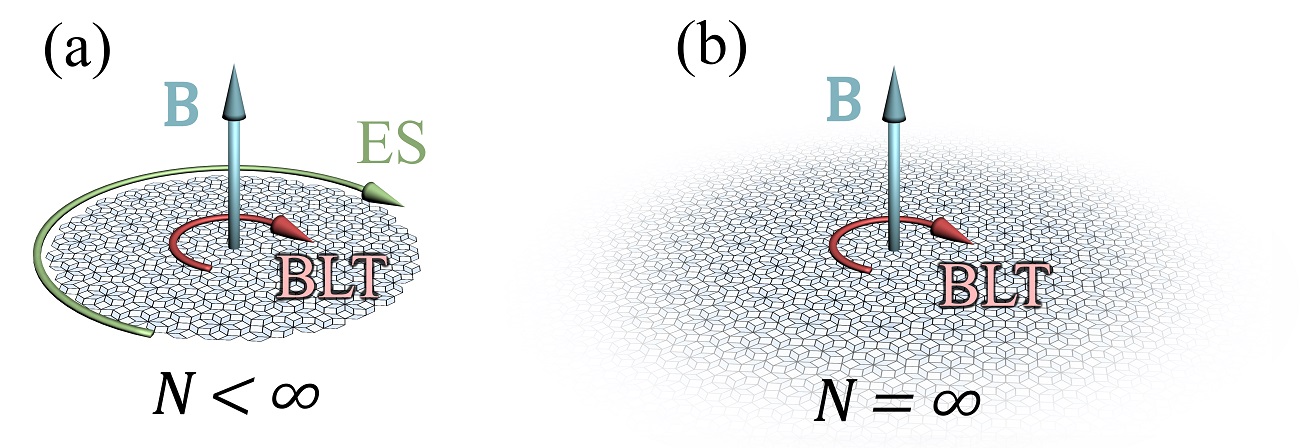}
	\caption{Illustration of the different types of in-gap states in (a) finite and (b) infinite quasicrystals when a uniform magnetic field (blue arrow) is present. In the case of a finite quasicrystal with $N$ sites (a), the green arrow depicts potential transport across a conventional ES, which forms at the boundaries of the lattice. The quasicrystal may also permit the formation of BLT, whose potential transport is depicted by the red arrow. In the case of a truly infinite quasicrystal (b), we no longer have boundaries and hence no ES, but we retain the BLT state and it's supported transport.}
	\label{fig:ExIntroFig}
\end{figure*}

\subsection{Motivation}

In crystalline materials, e.g. condensed matter or cold atoms in optical lattices, the standard picture according to band theory is that a system is either an insulator or metal \cite{Rosenberg,Wannier}. During the 1980s, this picture began to change with the discovery of topological states of matter \cite{Hasan2010,Bansil2016,asboth2016}. For example, topological Edge States (ESs) can occur when a charged particle in a crystal interacts with an external magnetic field \cite{Thouless1982,aidelsburger2015}. The physics of charged particles in a two-dimensional crystalline lattice with an applied strong magnetic field is a well-studied problem for both the single-particle \cite{Hofstadter1976,Aoki1996,Oktel2012,Harper2014,Oktel2017,Duncan2018,Du2018} and many-body \cite{Wang2014,Moller2015,tai2017,Gerster2017} regimes. There have also been numerous experimental realisations and proposals \cite{Dalibard2011,bloch2012,aidelsburger2015,dean2013,mukherjee2017}.

%Topological states of matter are usually characterised by topological invariants, which are themselves a property of the lattice bulk \cite{Hasan2010,Bansil2016,asboth2016}. Via the bulk-boundary correspondence, these invariants are related to the appearance of edge states that support transport. In particular, two-dimensional lattices subject to a magnetic field are characterised by the Chern number of each respective band \cite{Thouless1982,aidelsburger2015}. Differences between Chern numbers of the bands above and below a band gap are then equivalent to the number of Edge States (ESs) that appear within the band gap. To obtain such topological invariants, integrals are normally taken across the first Brillouin zone of the system \cite{Thouless1982,Wang2010}. This explicitly means that the underlying lattice needs to be periodic for these integrals to be generic properties of the bulk.

%The physics of charged particles in a two-dimensional periodic lattice with an applied magnetic field is a well-studied {and well-understood} problem, for both the single-particle \cite{Hofstadter1976,Aoki1996,Oktel2012,Harper2014,Oktel2017,Duncan2018,Du2018} and many-body \cite{Wang2014,Moller2015,tai2017,Gerster2017} regimes. There have also been numerous experimental realisations and proposals for the periodic case \cite{Dalibard2011,bloch2012,aidelsburger2015,dean2013,mukherjee2017}. However, alternative techniques must be employed to probe the in-gap states of a system when periodic boundary conditions cannot be consistently applied, e.g. in disordered or quasicrystalline systems.

Quasicrystals, on the other hand, are quasiperiodic structures with a long-range, self-similar nature \cite{Shechtman1984,Levine1984,Ishimasa1985,janssen2018,janssen2000}. This makes their features distinct from both periodic and disordered lattices. The general electronic properties of quasicrystals are little understood, especially in comparison to their periodic counterparts \cite{janssen2018,berger2000}. Two-dimensional quasicrystalline systems have been proposed and now experimentally realised to varying degrees in ultracold atoms \cite{Gopalakrishnan2013,Bordia2017,viebahn2018,Viebahn2019}, graphene bilayers \cite{Ahn2018}, and photonics \cite{freedman2006,Bandres2016,An2020}.

Recently, there has been renewed interest in adding a magnetic field to scenarios involving quasicrystalline lattices \cite{Hatakeyama1989,Tran2015,Fuchs2016,Fuchs2018,Huang2018,Huang2018b,Duncan2020}. In quasicrystals, the concepts of bands and band-gaps are difficult to consistently define, since Bloch's theorem is not enforceable without approximations to the overall structure.  While recent results have confirmed the presence of ESs in a magnetic field \cite{Tran2015,Duncan2020} and studied the appearance of higher-order topological states \cite{Chen2020,Hua2020,Varjas2019} in quasicrystals, there have been few tangible differences from their study in periodic systems.

In this paper, we will show that the now standard picture of insulators, metals and topological insulators with surface states is not the full story for quasiperiodic systems. When the confining potential is quasicrystalline there are two competing not necessarily commensurate length scales from the magnetic field and quasiperiodic lattice. This results in a magnetic aperiodicity which directly leads to the observation of Bulk Localised Transport (BLT) states. As BLT states arise from the magnetic aperiodicity, they are significantly different from previous states found in the internal sections of fractal lattices \cite{Marta2018,Pai2019,sarangi2021}. These fractal lattices are almost entirely composed of effective hard edges with no discernible bulk, meaning these `internal' ESs are truly just conventional ESs on an unconventional lattice. The BLT states are not an artefact of effective edges introduced through an impurity or set of dislocations \cite{ran2009,li2018,Duncan2018,valiente2019,Wang2020,Diop2020} from the quasicrystalline lattice. We will also note that the BLT states appear to be of a different character than the higher-order zero-dimensional corner modes found in quasicrystals \cite{Chen2020}, as these are entirely bound to corners on the hard boundary of a finite system. The BLT states do share many of the properties of ESs, but are entirely localised within the bulk, as illustrated in Fig.~\ref{fig:ExIntroFig}. Importantly, the BLT states support transport (or currents) in the same way as the well-studied ESs,  which could make them useful for future applications utilising BLT.

\subsection{Overview}

Before detailing the methods and calculations, we overview the main results of our work. This section is intended to review our work, with references to the rest of this paper where we discuss the approach and results in detail.

\emph{Quasicrystals and magnetic fields.} We envisage a system of charged particles existing on a quasicrystalline lattice under the influence of a uniform perpendicular magnetic field. This system is illustrated in Fig.~\ref{fig:ExIntroFig}. The Hamiltonian of this system is well described by the Hofstadter vertex model, introduced in Sec.~\ref{sec:Quaks}, which modifies the standard Hofstadter model to the vertex model of a quasicrystalline tiling, illustrated in Fig.~\ref{fig:IntroVertex}. Solving for the finite size spectrum and states of the Hofstadter vertex model is relatively straight forward. However, defining bands and in-gap states is difficult due to the breakdown of Bloch's theorem. We define the in-gap states, and hence the bands, via topological measures that are non-zero for in-gap states and outlined in Sec.~\ref{sec:MeasProp}.

\emph{Infinite size quasicrystals.} As the states we will consider are supported by the bulk of the lattice, it is interesting to consider if they are retained in the spectrum of the infinite Hamiltonian. To consider the infinite-size quasicrystalline lattice, we will utilise an infinite-size error controlled algorithm; which abandons square truncation to allow for the interaction of sites outside a finite patch of the lattice to be accounted for. The infinite-size algorithm is described in Sec.~\ref{sec:Infinite} and can be utilised to calculate the spectrum of other infinite-dimensional operators \cite{Colbrook2019}. We also detail a new extension of the infinite-size algorithm in order to calculate topological measures, as described in Sec.~\ref{sec:Top}. This allows for states with in-gap characteristics in the infinite-size quasicrystal (or any crystal or aperiodic lattice) to be identified.

\emph{BLT states.} We first illustrate BLT states in Sec.~\ref{sec:BLS} by considering in detail the Hofstadter vertex model of the quasicrystalline Ammann--Beenker (AB) tiling. As already stated, the BLT states are peculiar as they are in-gap, but entirely localised within the bulk of the lattice, with no component on the edge. Examples of BLT states are given for the AB tiling in Fig.~\ref{fig:BLS1}(c) and~\ref{fig:BLSfluxfix}, clearly illustrating the bulk nature of these in-gap states. The BLT states are also shown to be in-gap from their non-zero topological measures. Interestingly, the proportion of states that are of BLT type in a finite system converges to a non-zero value with increasing system size, as shown in Fig.~\ref{fig:SizeScale}. This is contrary to the regular ESs; which become a vanishingly small proportion of the states with increasing system size. As expected from the arguments illustrated in Fig.~\ref{fig:ExIntroFig}, we find that BLT states exist in the infinite-size spectrum, and are of the same character as those in the finite system, see Fig.~\ref{fig:BLSinf} for examples. We confirm that BLT states exist in the spectra of other quasicrystals for both finite and infinite cases in Sec.~\ref{sec:Other}. For example, we consider 5-fold, 7-fold, 10-fold, and 12-fold examples in Fig.~\ref{fig:BLSinfOther}. We also show examples of BLT states in Fig.~\ref{fig:BLSinfNSymm} where the rotational symmetry is broken and severe defects are included. In summary, we find that BLT states are ubiquitous in quasicrystals in magnetic fields, even in the presence of extreme deformations to the lattice.

\emph{Supported bulk transport.} Key to the future consideration and application of BLT states is the fact that they support transport like any other in-gap state. This results in BLT which is independent of the existence or form of any edges. In this way, BLT can be considered to be robust against perturbations along the edge, much like regular edge states are robust against perturbations in the bulk. We show explicitly that transport is supported by the BLT states of the AB Hofstadter vertex model in both the finite and infinite size cases in Figs.~\ref{fig:TransFinite} and~\ref{fig:TransInfinite} respectively. We also show that the location of this transport can be varied due to the BLT states being supported on different parts of the bulk. BLT will also exist in cases of extreme deformations to the system, as BLT states are retained. We also show in Fig.~\ref{fig:BLSinfOther_te} that the BLT states of other quasicrystals also support transport along them as would be expected. The BLT discussed and exhibited in this paper is not in direct correspondence with the regular transport of an electron in a magnetic field. On a lattice, the cyclic motion of the electron is truly in correspondence to the cyclic motion around a single tile of the lattice. BLT is a larger collective effect of the quasicrystalline lattice and magnetic field which goes beyond the circling of a single tile but it is still, of course, cyclic in nature.

\begin{figure}[t]
	\centering
	\includegraphics[width=0.75\linewidth]{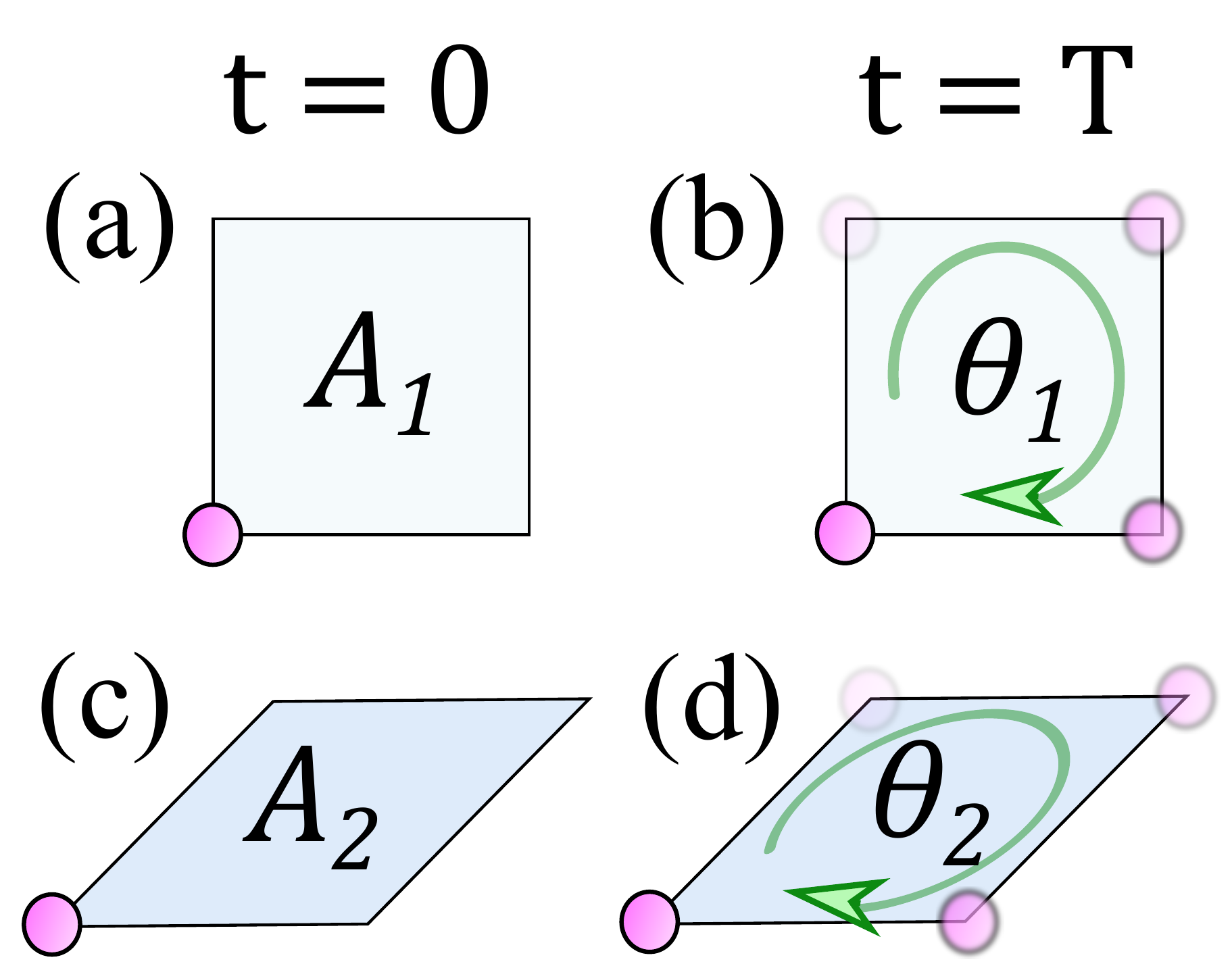}
	\caption{Origin of magnetic aperiodicity, showing two rhombic prototiles of a quasicrystalline tiling. In (a,c), a particle (denoted by the purple circle) is initially localised to one corner of the prototile. Each tile has a unique area of $A_1$ and $A_2$, and their fraction $A_1/A_2$ is irrational. After some time $T$ in (b,d), the particle circulates around the prototile and acquires a unique phase of either $\theta_1$ or $\theta_2$. The fraction $\theta_1/\theta_2$ will again be irrational due to the presence of incommensurate areas, leading to a magnetic aperiodicity on the full quasicrystalline tiling.}
	\label{fig:qcTiles}
\end{figure}

\emph{Origin of BLT states.} We find through example toy models in Sec.~\ref{sec:ToyModel} that the BLT states are a direct result of the interplay of the length scales between the magnetic field and quasicrystal. We coin the term magnetic aperiodicity in relation to these competing length scales. The origin of this interplay can be motivated by considering a particle looping around the individual distinct tiles of the quasicrystalline lattice, as illustrated in Fig.~\ref{fig:qcTiles}. Each distinct tile has a different area, which is incommensurate in relation to the areas of the other distinct tiles. When a particle circulates one of these tiles, it gains a phase that is dependent on the flux, which in turn is dependent on the area of the tile, as illustrated in Fig.~\ref{fig:qcTiles}. This means when a particle circulates around each distinct tile it gains a phase that is incommensurate with the phase associated with circulating the other distinct tiles, as the areas are incommensurate to each other and the magnetic field is uniform. It is the convolution (or interference) of these incommensurate phases that is the interplay of the length scales of the magnetic field and quasicrystalline lattice. With the end result being the generation of BLT states and support of BLT.

\subsection{Terminology}

As this work straddles the fields of condensed matter,  quantum simulators, and spectral computations from mathematical physics, there are certain terminologies we need to ensure are defined consistently to avoid confusion between readers of different fields. 

First, we will refer to a lattice as being periodic or crystalline as long as a unit cell can be defined with an associated Brillouin zone. This allows for the use of Bloch's theorem to calculate the band structure of the lattice. Note, this does not exclude the presence of edge states, as they are dependent on the boundary. Even in the case of open boundaries, bands can still be calculated. Therefore, a periodic lattice is defined independent of the boundary conditions, which would usually be considered as periodic, infinite or open. The strict definition of a lattice itself can be considered to only apply to periodic systems. However, the definition of a lattice in condensed matter physics is more general and can be interpreted as defining a group of discrete connected points. Throughout this work, we will follow this convention and refer to lattices as being any group of discrete connected points.

We will often refer to bulk states and in-gap states. Bulk states are all states that are allowed in the system with real quasimomentum according to Bloch's theorem. In-gap states are all other solutions to the Schr\"odinger equation for the lattice Hamiltonian, which are the complex quasimomentum solutions \cite{Duncan2018}. In general, we do not have access to the quasimomentum from numerical approaches. As Bloch's theorem breaks down in quasicrystals, we must turn to alternative methods to define if a state of the spectrum is in-gap. For this we turn to the topological measures of the Bott index and local Chern marker, as discussed in Sec.~\ref{sec:Top}. 

The meaning of an edge state must also be clearly defined. In spectral problems of infinite-size operators studied by the applied mathematics community, one of the main problems tends to be the removal of spectral pollution. Spectral pollution refers to a set of states in the spectrum which are not actually part of the infinite-size spectrum. These typically manifest in the form of edge states due to finite size effects. While the edge states do not exist in the spectrum of infinite-size operators, they are physical states of finite-size system, with distinct observable properties. Note, the BLT states outlined in this paper are not spectral pollution, and are in fact part of the spectrum of the infinite-size operator, which we will show in Sec.~\ref{sec:Zoo}. This means that not all in-gap states are spectral pollution, as is usually thought.

\begin{figure}[t]
	\centering
	\includegraphics[width=0.9\linewidth]{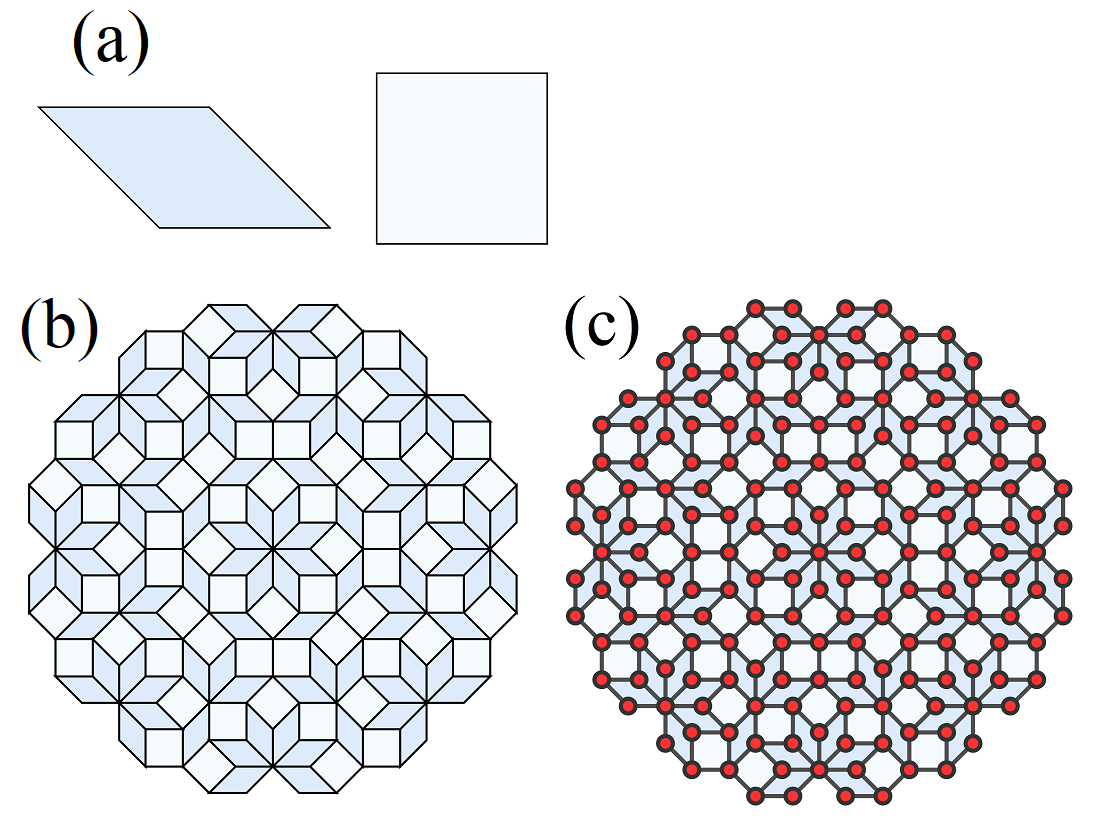}
	\caption{Construction of the Ammann--Beenker (AB) tiling, using the (a) incommensurate square and rhombus as prototiles. The aperiodic tiling is generated from these prototiles, leaving no gaps. Here, we take a circular cutoff in tiling space to show a (b) finite sample of AB tiling and preserve rotational symmetry with respect to the origin (centre of the tiling). Note, the enforcement of rotational symmetry here is arbitrary and plays no role in the formation of BLT in this study, as we show in Sec. \ref{sec:Other}.  The corresponding vertex model (c) is then defined by setting bonds as the edges of tiles, and the lattice sites as the intersection of tile edges. In this example, the total number of lattice sites is $N=185$.}
	\label{fig:IntroVertex}
\end{figure}

\section{Models of quasicrystals in magnetic fields}
\label{sec:Quaks}

\subsection{Hofstadter Vertex Model}

We will consider lattices generated from the vertex model of aperiodic tilings. A 2D tiling is a countable family of closed sets (prototiles) which covers the entire 2D plane without any gaps or overlaps \cite{grunbaum1987,janssen2018,Baake2002}. Aperiodic tilings are a subclass of tilings that exhibit long-range order, but no short-range translational invariance. Finding tiles that enforce quasiperiodicity is not a simple task, and the initial aperiodic tiling patterns contained thousands of distinct tiles \cite{janssen2018}. Penrose discovered an aperiodic tiling requiring only a few rhombic tiles \cite{penrose1974}. Since then, there has been a multitude of aperiodic tilings discovered with a variety of non-crystalline rotational symmetries \cite{janssen2018,cockayne2000}. We will focus on an AB vertex model as an example, which has 8-fold rotational symmetry and may be generated from an incommensurate rotation and projection of the 4D hypercubic lattice \cite{Socolar1989,lagarias1996,Maskov1998} (see Appendix \ref{app:Tiles}). We illustrate the quasicrystalline AB vertex model and its construction from an aperiodic tiling in Fig.~\ref{fig:IntroVertex}. However, our results are not specific to this 8-fold tiling, and we will show that BLT can occur in other quasicrystalline lattices in Sec.~\ref{sec:Other}.

The vertex model takes the aperiodic tiling and considers a lattice site to exist at each vertex and a bond to be present along the edges of tiles \cite{Rieth1995,Choy1985,Repetowicz1998,Prunel2002}, as shown in Fig.~\ref{fig:IntroVertex}(b) to (c). We will consider the vertex model of the AB tiling with a perpendicular constant magnetic field, as depicted in Fig.~\ref{fig:ExIntroFig}. The single-particle Hamiltonian is then
\begin{equation}
H = - J \sum_{\langle j,k \rangle}^N e^{i \theta_{jk}} \ket{j} \bra{k},
\label{eq:H}
\end{equation}
where $\theta_{jk}$ is the Peierls phase \cite{peierls1997} due to the magnetic field between sites $j$ and $k$, $\langle j,k \rangle$ is the sum over all $N$ vertices/sites connected by an edge, and $\ket{j}$ the state of a particle occupying site $j$. We consider the Landau gauge $\mathbf{A}(\mathbf{r}) = Bx \mathbf{\hat{y}} = (\phi/A)x \mathbf{\hat{y}}$, with the magnetic field strength $B$, flux $\phi$ (measured in terms of the flux quantum $\phi_0 = 2\pi$) and penetrating area $A$. We will take the area $A$ to be that of the square tile of the AB tiling (the qualitative results, including the observation of BLT states, are independent of the choice of $A$). Units of $\hbar=e=1$ are considered throughout this work, and we will work in units of energy $J$. The Hamiltonian~\eqref{eq:H} is well-understood when applied to periodic systems \cite{Hofstadter1976} and can even result in similar physics when applied to some quasicrystals \cite{Tran2015,Fuchs2016,Fuchs2018,Duncan2020}.

\subsection{Infinite-Size Algorithm}
\label{sec:Infinite}

\begin{figure}[t]
	\centering
	\makebox[0pt]{\includegraphics[width=0.48\textwidth]{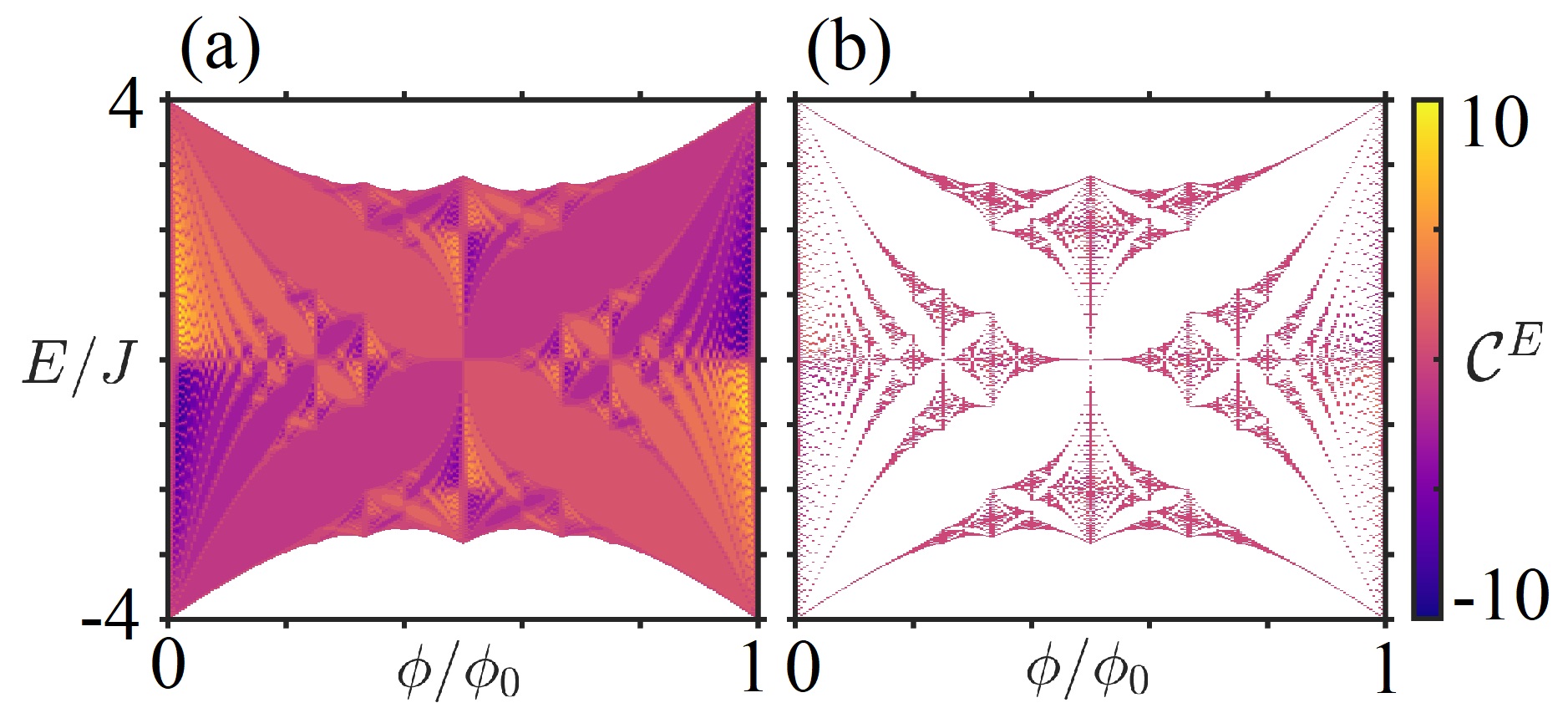}}
	\caption{Obtaining the Hofstadter butterfly for the infinite square lattice from the infinite-size algorithm discussed in Sec.~\ref{sec:Infinite}. We show the effective Chern marker, defined in Sec.~\ref{sec:Top}, for each state, showing the $\mathcal{C}^{E}$ (a) over a full range of energy values $E$, which includes states that are considered to be spectral pollution, and (b) over the infinite-size square lattice Hofstadter butterfly, with a restricted range of energy values which are permitted in the spectrum of the infinite lattice.}
	\label{fig:IntroInf}
\end{figure}

To rigorously probe the spectral properties of the infinite tiling directly, we will use a set of new computational techniques for infinite-dimensional spectral problems \cite{colbrook2020foundations}. We begin with a description of computing the spectrum, with further details presented in Appendix \ref{app:Inf}. As an example, the results of the infinite algorithm are shown for an infinite square lattice with a Hamiltonian of Eq.~\eqref{eq:H} in Fig.~\ref{fig:IntroInf}. The algorithm perfectly replicates the fractal Hofstadter butterfly usually generated through the consideration of periodic boundary conditions \cite{Hofstadter1976}. The removal of ESs, or in this case (of the infinite tile) spectral pollution, can be seen by the difference between Figs.~\ref{fig:IntroInf}(a) and (b). As discussed previously, spectral pollution is the terminology used to describe spurious eigenvalues present between parts of the essential spectrum.

We will utilise a new algorithm \cite{Colbrook2019}, developed by one of the co-authors, which allows for the calculation of the spectrum of a full infinite-dimensional operator with error control. This algorithm is general in its applications and will be of use to physical scenarios other than that considered here. For example, extensions to unbounded operators and partial differential operators can be found in Ref.~\cite{colbrook2019b}. For the present paper, the algorithm is of particular use since (i) the aperiodic nature of quasicrystals makes it a considerable challenge to approximate the spectrum of the full infinite-dimensional operator without finite-size effects and (ii) the approximation error can be computed and reaches effectively zero from a physical standpoint, as detailed below. There has also been an approach developed to obtain the exact solutions of quasicrystals of infinite size through the use of a superspace \cite{Valiente2021}. This method is specifically constructed to handle quasicrystalline problems by converting them to higher-dimensional periodic problems and could be applied in the scenarios discussed in this work. However, we find the algorithm utilised here to be highly efficient at handling the problem of a quasicrystal of infinite size in a uniform magnetic field.

In infinite dimensions, the Hamiltonian $H$ can be represented by an infinite Hermitian matrix, $\hat H=\{\hat H_{ij}\}_{i,j\in\mathbb{N}}$, which acts on $l^2(\mathbb{N})$, the space of square summable sequences. A suitable ordering of the sites (e.g. ordering by positional radius from an origin) leads to a matrix $\hat H$ with finitely many non-zero entries in each column. In other words, there exists a function $f:\mathbb{N}\rightarrow\mathbb{N}$ such that $\hat H_{ij}=0$ if $i>f(j)$, thus describing the sparsity of $\hat H$. Sparse Hamiltonians are a subclass of operators that are dealt with in Ref.~\cite{Colbrook2019} by considering the function
\begin{equation}
F_n(z):=\sigma_1(P_{f(n)}(\hat H-z)P_n),
\end{equation}
where $P_m$ denotes the orthogonal projection onto the linear span of the first $m$ basis vectors and $\sigma_1$ denotes the smallest singular value of the corresponding rectangular matrix. The rectangular truncation $P_{f(n)}(\hat H-z)P_n$ corresponds to including all of the interactions of the first $n$ sites (the first $n$ columns of $\hat H$) without needing to apply boundary conditions (see, for example, Fig. 1 of Ref.~\cite{colbrook_IMA_LT}). This is in sharp contrast to standard methods that typically take a square truncation of the matrix $\hat H$ (corresponding to a truncation of the tile) with a boundary condition. This difference allows us to prove convergence, provide error control, and also lends itself to adaptive computations of the full infinite-dimensional operator. Physically, $F_n(z)$ is the square-root of the ground state energy of the folded Hamiltonian $P_n(\hat H-z)^*(\hat H-z)P_n$. In Ref.~\cite{Colbrook2019}, it is shown that $F_n(z)$ converges down to the distance of $z$ to the spectrum of $H$ (uniformly on compact subsets of $\mathbb{C}$) as $n\rightarrow\infty$ which, together with a local optimisation routine, leads to the computation of the spectrum and approximate states with error control as $n\rightarrow\infty$. Algorithmic steps are provided in Appendix \ref{app:Inf}.

\section{Measures and Properties}
\label{sec:MeasProp}

\subsection{Topological Measures}
\label{sec:Top}

Probing topological invariants in quasicrystalline systems is difficult, due to the ill-defined Brillouin zone and the breakdown of Bloch's theorem. For two-dimensional lattices subject to an external perpendicular magnetic field, the topological invariant of each respective band is its Chern number \cite{Thouless1982,aidelsburger2015}. Differences between Chern numbers of the bands above and below a band gap are then equivalent to the number of ESs that appear within the band gap via the bulk-boundary correspondence. To obtain such topological invariants, integrals are normally taken across the first Brillouin zone of the system \cite{Thouless1982,Wang2010}. This explicitly means that the underlying lattice needs to be crystalline for these integrals to be generic properties of the bulk. However, there are measures that are independent of the boundary and have been shown to be equivalent to properties of the Chern number. These are the Bott index \cite{Toniolo2018} and the local Chern marker \cite{Bianco2011}, which we will utilise here.

\subsubsection{Finite Systems}

The Bott index is a spectral quantity defined for each individual state, related to the commutativity of two matrices \cite{Toniolo2018,Zeng2020}. It is defined for the $n$th eigenstate as
\begin{equation}
\mathcal{B}^{n} = \dfrac{1}{2\pi} \Imm \left\{ \Tr \left[\log \left( \hat{V}^{n}_x \hat{V}^{n}_y \hat{V}^{n \dagger}_x \hat{V}^{n \dagger}_y \right)\right]\right\},
\label{eq:Bott}
\end{equation}
with $\hat{V}_{x/y}^n$ being the projected position operators,
\begin{equation}
\hat{V}_x^{n} = \hat{Q}^{n} + \hat{P}^{n} \hat{U}_x \hat{P}^{n},
\end{equation}
\begin{equation}
\hat{V}_y^{n} = \hat{Q}^{n} + \hat{P}^{n} \hat{U}_y \hat{P}^{n},
\end{equation}
for the $n$th state cumulative projections
\begin{equation}
\hat{P}^{n} = \sum_{m=1}^{n} \ket{m}\bra{m}, \, \, \, \hat{Q}^{n} = \hat{\mathbb{I}} - \hat{P}^{n},
\label{eq:Proj}
\end{equation}
where $\ket{m}$ is the $m$th eigenstate (we follow the usual convention of listing states in order of increasing eigenvalues) and the unitary diagonal position operators are
\begin{equation}	\label{eq_Uxy}
\hat{U}_x = \exp \big( 2\pi i \hat{x}_S \big), \, \, \, \hat{U}_y = \exp \big( 2\pi i \hat{y}_S \big).
\end{equation}
Note, the $\hat{x}_S$/$\hat{y}_S$ are position operators which must be scaled between 0 and 1. 

By calculating the Bott index, we can measure obstructions to the formation of a maximally localised Wannier basis that span the occupied states \cite{Loring2010,Wang2020b}. In order to find localised Wannier states, it is usually necessary to find continuous and periodic logarithms of families of unitaries \cite{cornean2016}. This is discussed in detail in Ref.~\cite{Cances2017}, where they also show that such logarithms can not be defined when the Chern numbers associated to the occupied states are non-zero. In essence the problem comes down to the definition of the fractional power of a matrix, and issues around the choice of a branch cut in the complex plane to define the required logarithm \cite{Cances2017,panati2007}. Note, this logarithm is not the same as the one utilised in the definition of the Bott index. If at a given energy the Bott index is non-zero, then any state with that energy must be in-gap, as the occupied band of states below must have a non-zero Chern number associated to it. The Bott index has proven useful in disordered and quasicrystalline systems, where bands cannot be defined through Bloch's theorem \cite{Huang2018,Huang2018b,Duncan2020,Wang2020b}. However, it is computationally expensive, as it requires the logarithm of a matrix whose size scales with the lattice. The Bott index does not lend itself to the large system sizes of this work and is, in general, ill-defined in the infinite-size case. Therefore, we will for the most part turn to an alternative measure in the local Chern marker. 

The topological invariant can also be projected into the real physical space of the system, and for the Chern number this is known as the local Chern marker \cite{Bianco2011,Song2019,caio2019,Irsigler2019}. Unlike the Bott index, the local Chern markers are defined on every single site $j$ of the lattice for the $n$th eigenstate as
\begin{equation}
C_j^{n} = -\frac{4\pi}{A_c^2} \mathrm{Im}\left\{ \bra{j} \hat{x}^{n} \hat{y}^{n} \ket{j} \right\},
\end{equation}
with $A_c$ a reference area of the lattice and 
\begin{equation}
\hat{x}^{n} = \hat{Q}^{n}\hat{x}\hat{P}^{n}, \: \: \hat{y}^{n} = \hat{P}^{n}\hat{y}\hat{Q}^{n},
\end{equation}
where $\hat{x}$ / $\hat{y}$ are the position operators. We will take $A_c$ to be the area of the square tile of the AB lattice. The local Chern marker has already been used to distinguish topological states in quasicrystals and disordered systems \cite{Tran2015,Kuno2019}. For large lattice sizes, we find that the local Chern marker is a more efficient way of distinguishing in-gap states since we do not need to compute a matrix logarithm for each state. Moreover, it can be extended to infinite systems. We will consider an effective Chern marker $\mathcal{C}^{n}$ for a given state $\ket{n}$, which takes the integer of maximal counts in the distribution of $C_j^{n}$, which we find to be in agreement with the Bott index for crystals and quasicrystals.

\subsubsection{Infinite Systems}

Associated with the Hamiltonian $H$ is a projection-valued measure, $\mathcal{E}$, whose existence is guaranteed by the spectral theorem \cite{reed1980} and whose support is the spectrum $\mathrm{Sp}(H)$. This diagonalises $H$, even when there does not exist a basis of normalisable eigenfunctions (recall that we are working in an infinite-dimensional Hilbert space):
\begin{equation}
 H=\int_{\mathrm{Sp}(H)} \lambda d\mathcal{E}(\lambda).
\end{equation}
In finite dimensions and for compact Hamiltonians, $\mathcal{E}$ consists of a sum of Dirac measures, located at the eigenvalues, whose values are the corresponding projections onto eigenspaces. More generally, however, there may be a continuous component of the spectrum and spectral measure. Generalisations of the spectral projectors in Eq.~\eqref{eq:Proj} can be given in terms of $\mathcal{E}$ as
\begin{equation}
\begin{aligned}
\hat{P}^{E} = \int_{(-\infty,E]} d\mathcal{E}(\lambda),
\end{aligned}
\label{eq:Proj244}
\end{equation}
where we now label over energy values $E$, which also covers the possibility of continuous spectra.

\begin{figure*}[t]
	\centering
	\makebox[0pt]{\includegraphics[width=0.7\textwidth]{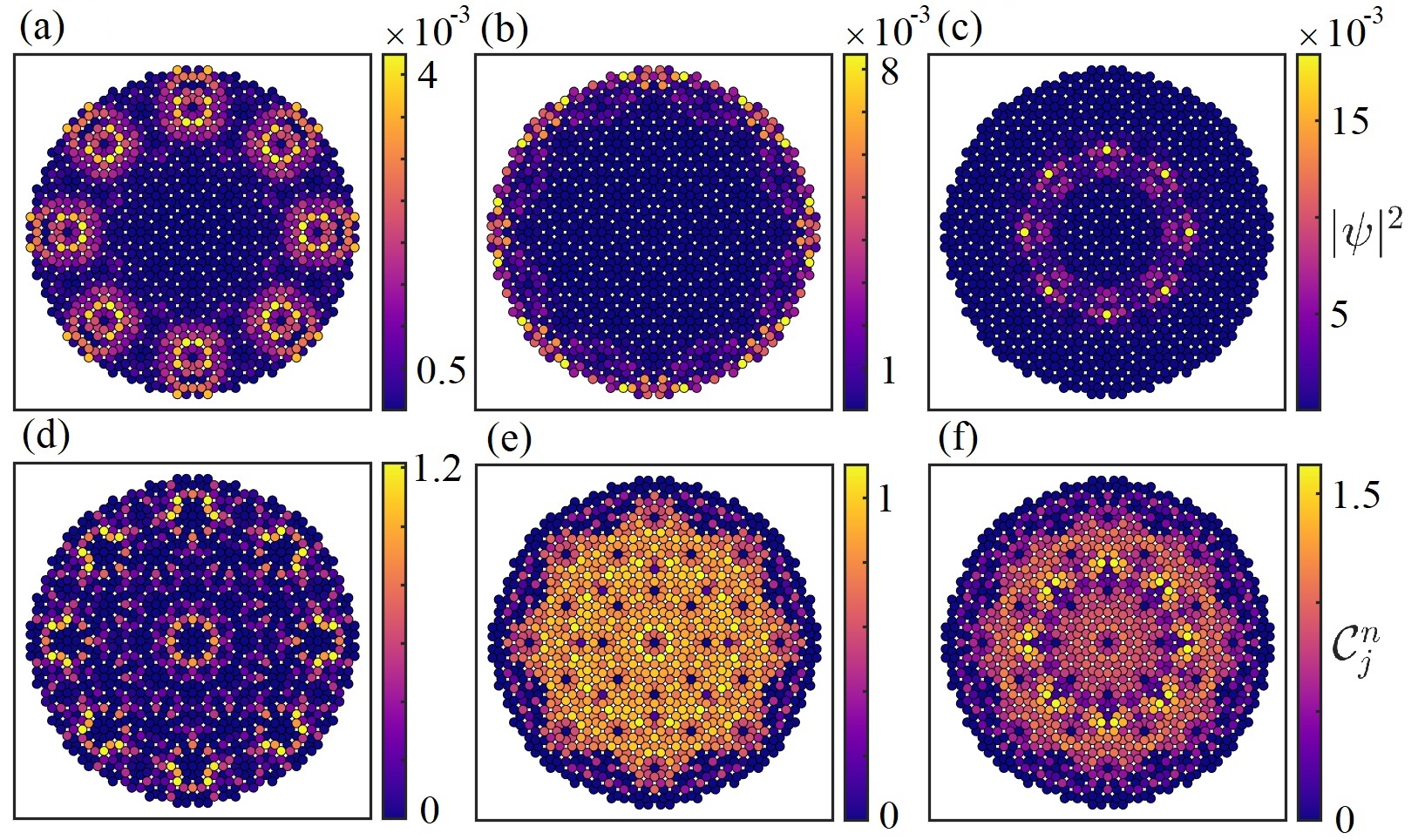}}
	\caption{Example states (a-c) and their $\mathcal{C}_j^{n}$ distributions (d-f), showing a (a,d) normal bulk eigenstate at $n=300$, (b,e) ES at $n=485$ and (c,f) BLT state at $n=491$. For each case, the corresponding Bott indices are (a) $\mathcal{B}=0$, (b) $\mathcal{B}=1$ and (c) $\mathcal{B}=1$. The $\mathcal{C}_i^{n}$ distributions have the minimum value saturated to $0$ for visual clarity.}
	\label{fig:BLS1}
\end{figure*}

The key ingredient that allows approximations of $\mathcal{E}$ to be computed is the formula for the resolvent
\begin{equation}
( H-z)^{-1}=\int_{\mathrm{Sp}( H)} \frac{d\mathcal{E}(\lambda)}{\lambda-z} .
\end{equation}
In Ref.~\cite{colbrook2019computing}, it is shown how to compute the action of the resolvent with error control via the rectangular truncations $P_{f(n)}(\hat H-z)P_n$. Using this, we compute a smoothed approximation of $\mathcal{E}$ via convolution with a rational kernel $K_\epsilon$ for smoothing parameter $\epsilon>0$. Taking $z=x+i\epsilon$, the classical example of this is Stone's formula which corresponds to convolution with the Poisson kernel
\begin{equation*}
\frac{1}{2\pi i}\left[(H-z)^{-1}-( H-\overline{z})^{-1}\right]=\int_{\mathbb{R}}\frac{\epsilon\pi^{-1}}{(x-\lambda)^2+\epsilon^2}d\mathcal{E}(\lambda).
\end{equation*}
As $\epsilon\downarrow 0$, this approximation converges weakly (in the sense of measures) to $\mathcal{E}$. However, for a given truncation size, if $\epsilon$ is too small the approximation becomes unstable due to approximating the sum of Dirac masses that correspond to the spectral measure of the truncation of $\hat H$. There is an increased computational cost for smaller $\epsilon$, which typically requires larger truncation parameters. Since we want to approximate spectral properties without finite-size effects, it is advantageous to replace the Poisson kernel with higher-order rational kernels developed in Ref.~\cite{colbrook2020}. This allows a larger $\epsilon$ for a given accuracy, thus leading to a lower computational burden. The reason for using rational kernels is that, through a weighted distribution of resolvents, we can recover a generalised Stone's formula
\begin{equation*}
[K_{\epsilon}*\mathcal{E}](x)=\frac{-1}{2\pi i}\sum_{j=1}^{m}\left[\alpha_j (H-(x-\epsilon a_j))^{-1}-c.c.\right],
\end{equation*}
which converges with $m$th order of convergence in $\epsilon$ \cite{colbrook2020}. Here, $c.c.$ denotes taking the adjoint, the constants $\alpha_j$ and $a_j$ can be found in Appendix~\ref{app:Inf}, and $*$ represents convolution. With this in hand, and for a given energy value $E$, we can write down smoothed generalisations of the spectral projectors in Eq.~\eqref{eq:Proj} as
\begin{equation}
\begin{aligned}
\hat{P}^{E}_{\epsilon} = \int_{-\infty}^E [K_\epsilon*\mathcal{E}](\lambda)d\lambda, \, \, \, \hat{Q}^{E}_{\epsilon} = \mathbb{I} - \hat{P}^{E}_{\epsilon},
\end{aligned}
\label{eq:Proj2}
\end{equation}
where $\mathbb{I}$ denotes the identity operator. Note that the convolution $[K_\epsilon*\mathcal{E}]$ is a \textit{bona fide} operator-valued function, and so the above definition makes sense. Finally, we define
\begin{equation}
\begin{aligned}
\hat{x}_{\epsilon}^{E} = \hat{Q}^{E}_{\epsilon} \hat{x} \hat{P}^{E}_{\epsilon}, \, \, \, \hat{y}_{\epsilon}^{E} = \hat{P}^{E}_{\epsilon} \hat{y} \hat{Q}^{E}_{\epsilon},
\end{aligned}
\end{equation}
and the smoothed infinite-dimensional local Chern marker on a basis site $j$ up to energy value $E$ as
\begin{equation}
\begin{aligned}
C_j^{E} = \dfrac{-4\pi}{A_c^2} \Imm \left\{ \bra{j} \hat{x}_{\epsilon}^{E} \hat{y}_{\epsilon}^{E} \ket{j}\right\}.
\end{aligned}
\end{equation}
In the examples that follow, we will take $\epsilon=0.05$ and the $6$th order kernel described in Appendix~\ref{app:Inf}, where we also give further algorithmic details. We will consider the same definition of an effective Chern marker, $\mathcal{C}^{n}$, for the infinite size as we do for the finite size.

As an example of this new infinite size algorithm for a topological measure, the results for the effective Chern marker for infinite size are shown for a square crystalline lattice with a Hamiltonian of Eq.~\eqref{eq:H} in Fig.~\ref{fig:IntroInf}. The Hofstadter butterfly has no states with a non-zero effective Chern marker present (as expected), apart from a handful of states which, for a given resolution, have a vanishing Chern marker as the algorithm converges. The algorithm developed here to compute topological properties for infinite size systems is of wide applicability and could be used to study other quasiperiodic, aperiodic, or even periodic lattices with complex structure.

\subsection{Radial Measure}

To distinguish between BLT states and ESs in the gap, we will define a radial measure of
\begin{equation}
\mathcal{L}^n = \frac{1}{N_f} \sum_{j} \rho^n_j r_j,
\label{eq:local}
\end{equation}
with $\rho^n_j = |\psi_j^n|^2/\mathrm{max}\left( |\psi^n|^2\right)$ being the rescaled probability density if this quantity is at least $0.75\mathrm{max}\left( |\psi^n|^2\right)$, and $\rho^n_j = 0$ otherwise. Here, $\psi^n$ is the wavefunction of the $n$th eigenstate, $N_f$ is the number of elements for which  $\rho^n_j>0$, and $r_j$ is the $j$th site normalised radial coordinate with $0 \leq r_j \leq 1$. Therefore, $\mathcal{L}^n $ is defined such that for every state $0 \leq \mathcal{L}^n \leq 1$, with ESs having $\mathcal{L}^n \sim 1$ and BLT states $\mathcal{L}^n < 1$. The radial measure gives the degree of which the density profile is localised towards the lattice centre. Note that, for regular bulk states, we also have that $\mathcal{L}^n < 1$. However, we distinguish between bulk states and in-gap BLT states via the topological measures already discussed.

\subsection{Transport Properties}

The most significant physics that emerges due to the presence of in-gap states is their resulting transport properties. For ESs, this means that transport is supported along the edge of the system. The BLT states characterised in this paper support transport along localised regions within the bulk of the lattice. For the finite lattice, this requires the evolution of the current state $\psi(t_0)$ under the time evolution such that the final state is
\begin{equation}
\psi(t_1) = e^{-i H (t_1-t_0)} \psi(t_0).
\end{equation}
In our calculations, we will use a Trotter decomposition of the evolution unitary into discrete time steps. Note, that our Hamiltonian is always time-independent and we do not drive the system in any way.

For the infinite size lattice, we cannot just apply the time evolution to a finite truncation, since we want to avoid finite-size effects. For a holomorphic function $g$, Cauchy's integral formula yields
\begin{equation}
g(H)=\frac{1}{2\pi i}\int_{\gamma}g(z)(H-z)^{-1}dz,
\end{equation}
where $\gamma$ is a closed contour looping once around the spectrum. Transport properties are computed via the choice $g(z)=\exp(-i zt)$. The contour integral is computed using quadrature and approximations of the resolvent $(H-z)^{-1}$ via rectangular truncations as above (see Ref.~\cite{colbrook2019computing} for details and an example of fractional diffusion on a quasicrystal). In particular, the rectangular truncation of the Hamiltonian is chosen adaptively through \textit{a posteriori} error bounds. This allows us to perform rigorous computations with error control that are guaranteed to be free from finite-size or truncation/discretisation effects, directly probing the transport properties of the infinite lattice. This is difficult to achieve via other methods such as truncating the tile since it can be difficult to predict how large the truncation needs to be \textit{a priori}.

\begin{figure}
	\centering
	\makebox[0pt]{\includegraphics[width=0.49\textwidth]{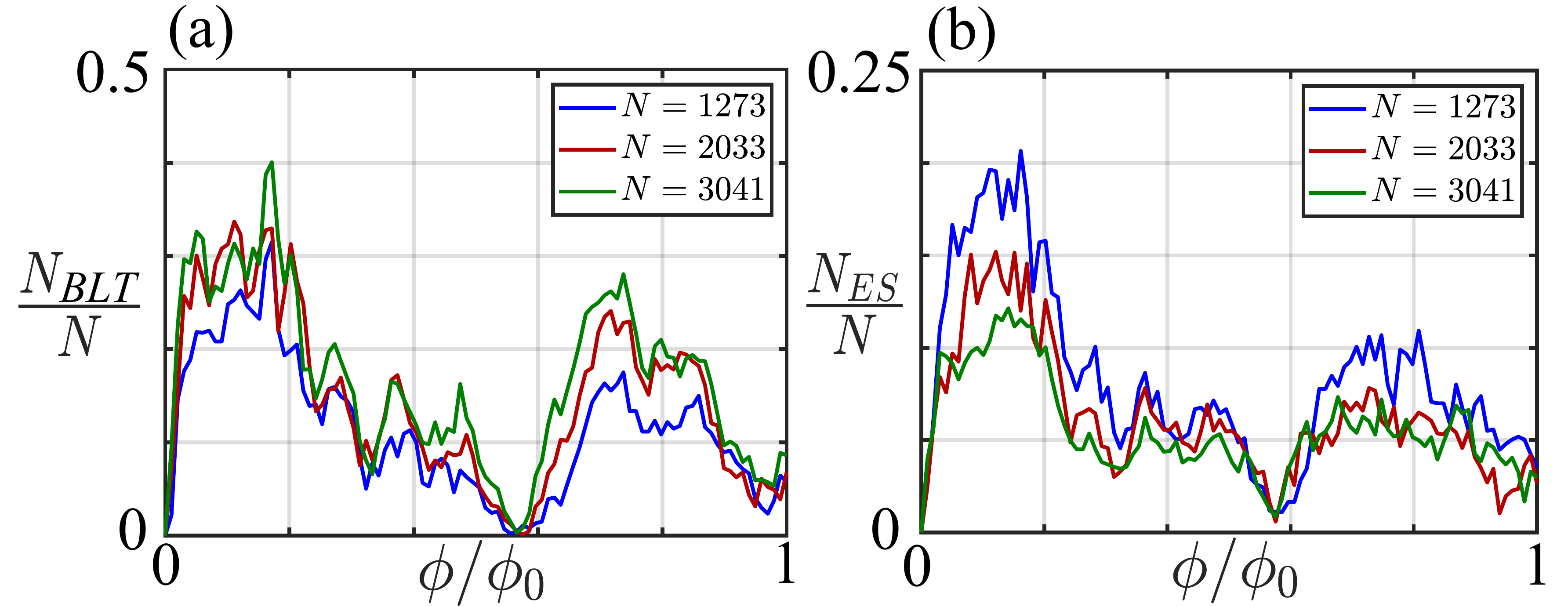}}
	\caption{Number of in-gap states as a function of flux, showing (a) the number of BLT states ($N_{BLT}$) and (b) the number of ESs ($N_{ES}$). The blue, red and green curves correspond to $N=1273$, $N=2033$ and $N=3041$ lattice sites respectively. Generally speaking, ESs usually fall in presence for the larger system sizes, as we expect. However, the number of BLT states may actually increase in some intervals of flux values.}
	\label{fig:FluxScale}
\end{figure}

\begin{figure}[t]
	\centering
	\makebox[0pt]{\includegraphics[width=0.49\textwidth]{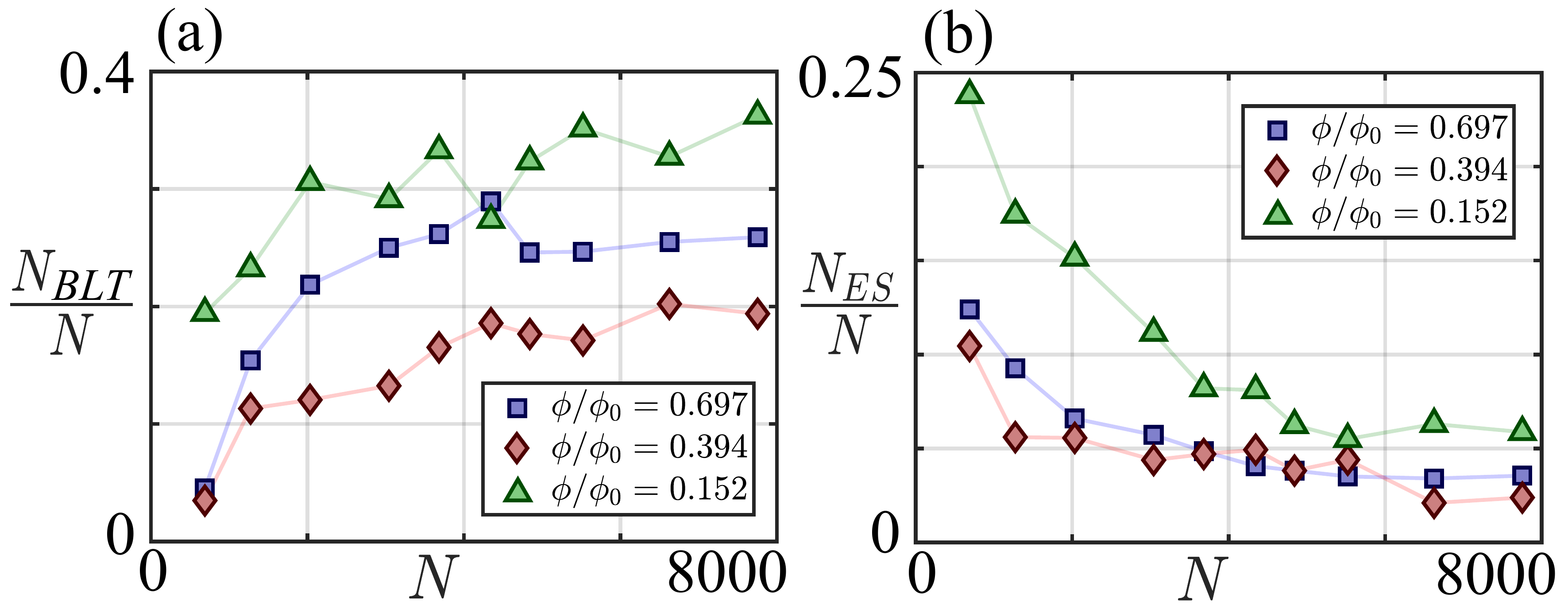}}
	\caption{Number of in-gap states as a function of system size, showing (a) the number of BLT states ($N_{BLT}$) and (b) the number of ESs ($N_{ES}$). The blue squares, red diamonds and green triangles correspond to a $\phi/\phi_0$ of $0.697$, $0.394$ and $0.152$ respectively. For each flux, there are signs of overall convergence at large $N$ for the total number of different topological states, but fluctuations can frequently occur due to the inhomogeneous nature of the lattice.}
	\label{fig:SizeScale}
\end{figure}

\begin{figure*}
	\centering,
	\makebox[0pt]{\includegraphics[width=0.9\linewidth]{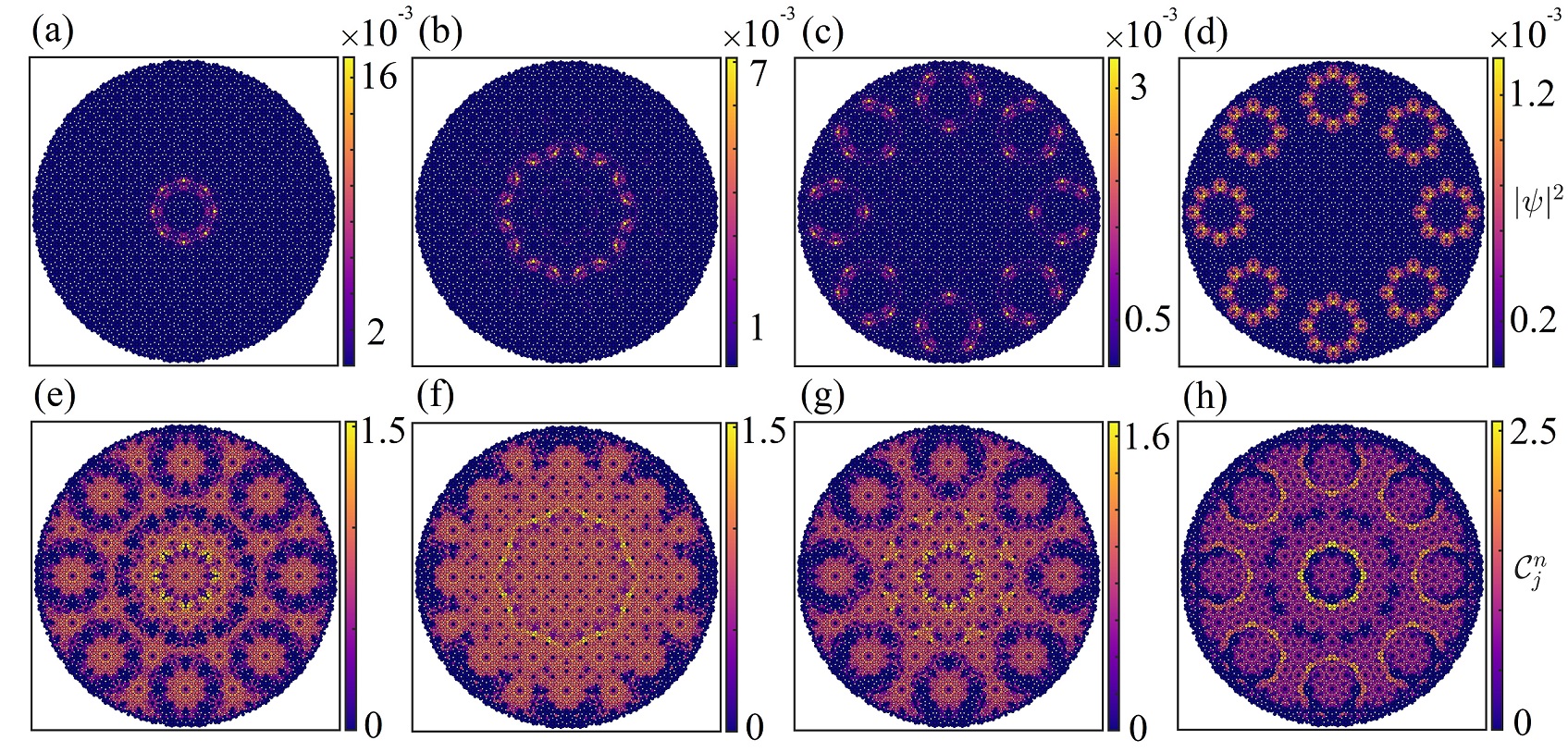}}
	\caption{Example BLT states (a-d) and their $\mathcal{C}_j^{n}$ distributions (e-h) for a larger portion of the AB tiling with $7753$ sites. Each state corresponds to (a,e) $n=3022$, (b,f) $n=2987$, (c,g) $n=3030$ and (d,h) $n=2219$.}
	\label{fig:BLSfluxfix}
\end{figure*}

\section{Bulk Localised Transport states}
\label{sec:BLS}

As already discussed, for Hamiltonian~\eqref{eq:H} on a crystalline lattice with open boundary conditions, there are in general two states that can be found: ordinary bulk states and in-gap ESs. In quasicrystals composed of single tiles, such as the Rauzy tiling, it is also observed that there are again two states, bulk states and in-gap ESs \cite{Tran2015}. For Hamiltonian~\eqref{eq:H}, this is all that would usually be expected unless there were perturbations or defects present. These perturbations could include, for example, the introduction of impurities, which can have in-gap states bound to them \cite{ran2009,li2018,Duncan2018,valiente2019,Wang2020,Diop2020}, or the presence of internal hard edges, as in fractal lattices \cite{Marta2018,Pai2019,sarangi2021}.

In quasicrystalline lattices constructed from multiple incommensurate tiles, a single peculiar in-gap bulk state was recently observed in Ref.~\cite{Duncan2020}. However, the origins of this state were not known and it was introduced as a potential one-off peculiarity. In this section, we show that the state observed in Ref.~\cite{Duncan2020} turns out to be one of many BLT states.

Examples of all three possible states, i.e. a bulk state, ES and BLT state are shown in Fig.~\ref{fig:BLS1} for a finite-size AB tiling vertex model with $1273$ sites, along with their corresponding local Chern marker. Note, in this figure, and other figures shown later, we saturate the colour maps of the local  Chern marker distributions to a range that shows the variation in values within the bulk. This is necessary because of large divergences in the local Chern marker near edges of the system, which occur in order for the sum of all local Chern markers in a given state to be zero \cite{Bianco2011,Tran2015}. In the bulk of the lattice, these fluctuations will generally be small, allowing for the effective Chern marker of a state to be visualised under a suitable range for the colourmap.

%We also consider some toy models for the formation of the BLSs in Sec.~\ref{sec:ToyModel}, where we find that it is the aperiodicity introduced through the interplay of the magnetic field with the incommensurate areas of the tiling that leads to the formation of BLSs. First, we will describe the scaling, form, and prevalence of the BLSs in the 8-fold AB lattice.

\begin{figure*}
	\centering
	\makebox[0pt]{\includegraphics[width=0.9\linewidth]{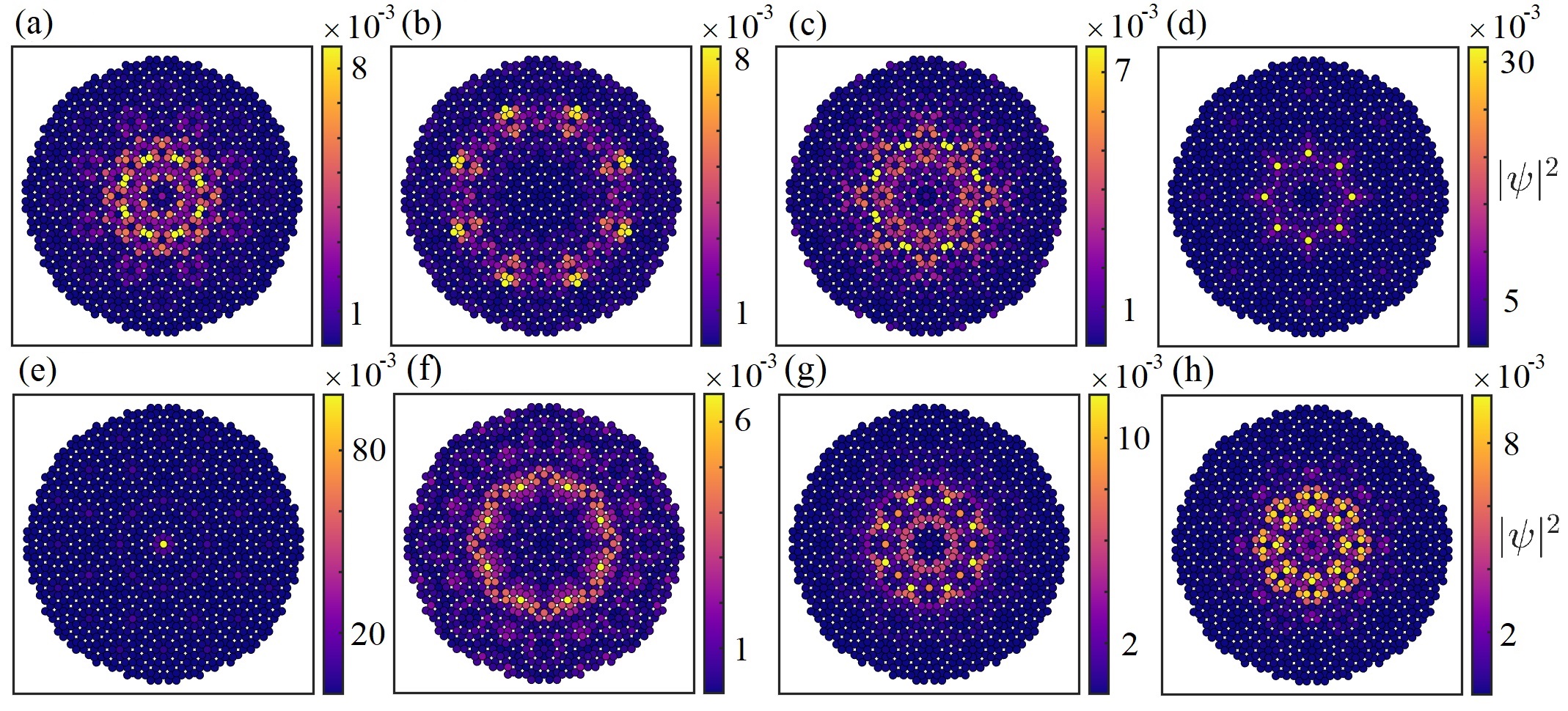}}
	\caption{Example BLT states at different flux. Each state corresponds to (a) $n=290$ at $\phi=0.141\phi_0$, (b) $n=379$ at $\phi=0.394\phi_0$, (c) $n=582$ at $\phi=0.546\phi_0$, (d) $n=90$ at $\phi=0.909\phi_0$, (e) $n=98$ at $\phi=0.909\phi_0$, (f) $n=291$ at $\phi=1.250\phi_0$, (g) $n=135$ at $\phi=1.579\phi_0$ and (h) $n=309$ at $\phi=2.623\phi_0$. For each case, the corresponding Bott indices are (a) $\mathcal{B}=-2$, (b) $\mathcal{B}=-1$, (c) $\mathcal{B}=-1$, (d) $\mathcal{B}=1$, (e) $\mathcal{B}=1$, (f) $\mathcal{B}=-1$, (g) $\mathcal{B}=1$, and (h) $\mathcal{B}=1$, meaning each state shown is in-gap. }
	\label{fig:BLSdiffflux}
\end{figure*}

\subsection{Scaling of the In-Gap States}

First, we split our spectra into the ES, BLT state and bulk state components. We split the spectra by using the radial measure of Eq.~\eqref{eq:local}, combined with the effective Chern marker of the state. We can see that if we vary the flux, as shown in Fig.~\ref{fig:FluxScale}, then the number of ESs or BLT states at any given flux can vary. There is also a characteristic dip in the in-gap states at a single flux within $0 \leq \phi/\phi_0 \leq 1$ as is the case for ESs in a crystalline lattice.

There is an interesting trend in the BLT states compared to the ESs, in that the number of BLT states as a proportion of the total number of states is increasing with system size. This is contrary to what is expected for normal ESs in the gap, which will decrease as a proportion of the spectrum with system size, due to the area of the bulk increasing in size faster than the edge increases in linear size. By considering larger system sizes for the finite system in Fig.~\ref{fig:SizeScale}, we show that, as expected, the proportion of ESs tends towards zero. However, the proportion of BLT states increases with system size and appears to converge towards a non-zero number. In fact, the values near convergence are $\sim 20 - 40\%$ of the total states, which is considerable and seen across a broad range of flux values. 

\begin{figure*}
	\centering
	\makebox[0pt]{\includegraphics[width=0.95\linewidth]{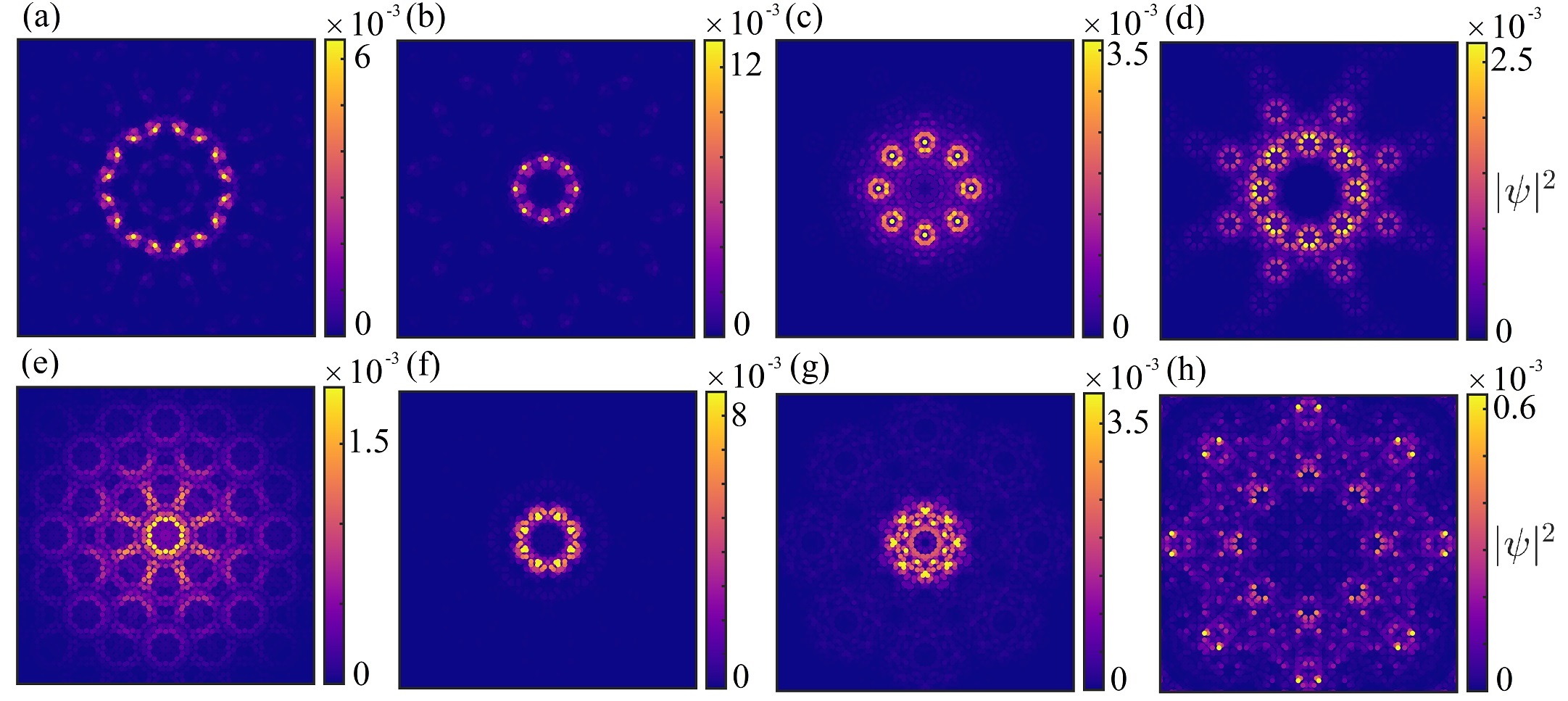}}
	\caption{Density profiles of example BLT states on the infinite tiling for (a,b) $\phi=0.69 \phi_0$, (c) $\phi=0.2 \phi_0$, (d,e) $\phi=0.4 \phi_0$ (f) $\phi=0.8 \phi_0$ and (g,h) $\phi=0.1 \phi_0$. Each state corresponds to an energy value (with shown error bounds) of (a) $E=-0.98730\pm 3\times 10^{-5}J$, (b) $E=-0.97479\pm 2\times 10^{-5}J$, (c) $E=-0.97743\pm 2\times 10^{-5}J$, (d) $E=-1.562934 \pm 8 \times 10^{-6}J$, (e) $E=-0.78855\pm 2 \times 10^{-5}J$, (f) $E=-0.997183\pm 5 \times 10^{-6}J$, (g) $E=-1.60422\pm 2\times 10^{-5}J$ and (h) $E=-1.5002\pm3 \times 10^{-4}J$. All states shown have a non-zero effective Chern marker, and are therefore BLT states with in-gap characteristics.}
	\label{fig:BLSinf}
\end{figure*}

The fact that BLT states do not make up a small percentage of the total states is a sign of their truly bulk nature. These are states that scale with the size of the bulk, rather than the edge. It is then reasonable to expect that their location in the lattice and density configuration will be as varied as that of the ordinary bulk states themselves. In other words, they can support transport over large and varied sections of the lattice bulk. This is in stark contrast to the transport supported by ESs, which is necessarily located along boundaries. It is important to mention here that the observed convergence with system size is no guarantee that the BLT states will be preserved for the infinite size lattice, as the usual thermodynamic limit approach to the infinite size is ill-advised in quasicrystals due to their aperiodic nature. We can then ask an intriguing question -- are the BLT states preserved for the infinite-size quasicrystal?

\subsection{A Zoo of Bulk Localised Transport States}
\label{sec:Zoo}

We can now report that there is in fact a rich and varied zoo of BLT states for the AB vertex model in a magnetic field, for both finite and infinite systems. The BLT states are also prevalent in a variety of other quasicrystals, which we will discuss in Sec.~\ref{sec:Other}.

We first fix the flux to $\phi/\phi_0 = 0.69$ and give examples of the varied structure of BLT states in Fig.~\ref{fig:BLSfluxfix}. To find the large variety of BLT states realisable at this flux, we simply need to extend the system size from the previous consideration in Fig.~\ref{fig:BLS1}. The examples shown in Fig.~\ref{fig:BLSfluxfix} are for $7753$ sites in the 8-fold lattice. We show that not only is the original BLT state retained, as shown in Fig.~\ref{fig:BLSfluxfix}(a), but we also realise BLT states in different regions of the lattice, both far into the bulk and nearer the boundary of the system. The form of the states shown in Fig.~\ref{fig:BLSfluxfix} further explains the trend of the proportion of BLT states as we increase the system size shown in Fig.~\ref{fig:SizeScale}. As the quasicrystal becomes larger, there are more regions of the lattice to which the BLT states can localise to. We also show the local Chern markers for each state in Fig.~\ref{fig:BLSfluxfix}, where it can be seen that the effective Chern marker of the bulk is non-zero, as the majority of sites in the bulk have a non-zero local Chern marker.

BLT states are not a peculiarity of a single flux, as was shown in Fig.~\ref{fig:FluxScale}, and we show a range of example BLT states at different flux in Fig.~\ref{fig:BLSdiffflux}. All the states shown in Fig.~\ref{fig:BLSdiffflux} have a non-zero Bott index and effective Chern marker. It is clear from the examples of Fig.~\ref{fig:BLSdiffflux} that the BLT states are not an artefact of a single region or a subset of regions of the lattice. Instead, they appear throughout the system, with their location dependent on the flux. This hints that their origins are due to an interplay of the constant magnetic field with the quasiperiodicity of the lattice, which we will explore further in Sec.~\ref{sec:ToyModel}. The appearance of BLT states is as diverse as the usual bulk states observed on the quasicrystal, reflecting the quasiperiodic nature of the system. In Fig.~\ref{fig:BLSdiffflux}(d,e), examples are shown which have localisation to a single site, or a rotationally symmetric set of single sites, with other examples showing intricate localised structures. A key property of the BLT states appears to be their ability to appear anywhere in the quasicrystalline lattice. Their ability to be localised at differing regions of the lattice could prove useful in future applications, as their position is not restricted and could be tuned without altering the lattice geometry.

The truly intriguing question for BLT states is whether they survive in the infinite-size system? We confirm that the BLT states can indeed exist in the infinite-size quasicrystal by applying the method of Sec.~\ref{sec:Infinite}. A set of example states are shown in Fig.~\ref{fig:BLSinf}, all with a non-zero effective Chern marker. The similarity of these BLT states to the finite-size results already discussed is striking. Again, it is also shown that the region where BLT states localise is flux-dependent. The persistence of BLT states into the infinite-size is an important difference for the usual edge states of Hamiltonian~\eqref{eq:H}. From periodic systems, we would expect all states in the infinite size to be in the bulk bands, with no in-gap states. However, with the BLT states preserved, a quasicrystal in a magnetic field can have states with properties that are usually considered to be related to being in-gap, even for the infinite-size lattice without defects or boundaries.

\subsection{Prevalence of the Bulk Localised Transport States}

\begin{figure}
	\centering
	\makebox[0pt]{\includegraphics[width=0.42\textwidth]{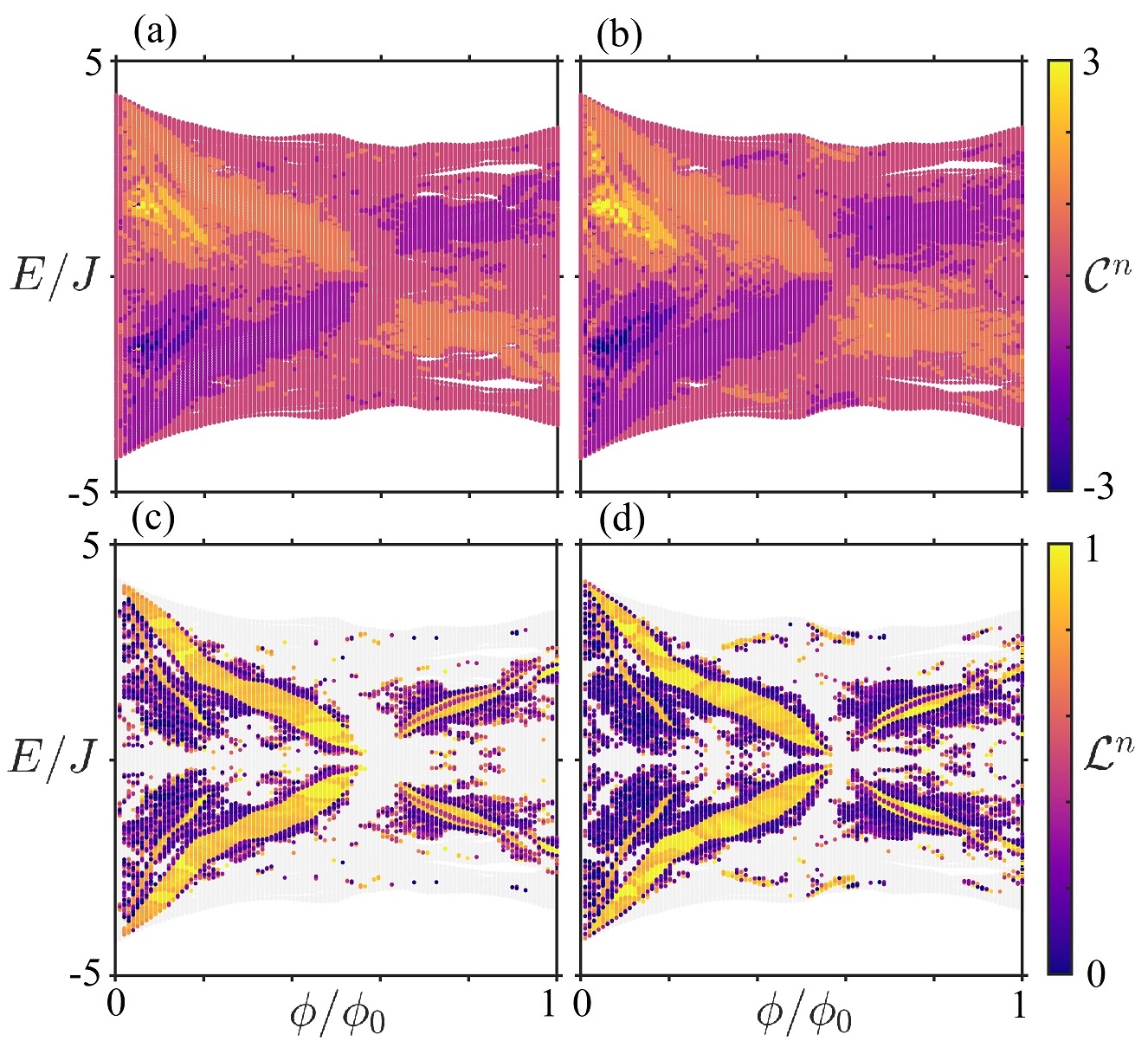}}
	\caption{Spectra of the quasicrystalline lattice, where each state is labelled according to the effective Chern marker $\mathcal{C}^{n}$ and radial measure $\mathcal{L}^{n}$ for (a,c) $N=1273$ and (b,d) $N=3041$ sites. The spectra (a) has similar properties to what is observed in the square lattice, but with a more intricate and quasiperiodic structure. For the larger lattice (b), one also observes that the total number of topological states throughout the spectra increases. For visualisation purposes, states with zero $\mathcal{C}^{n}$ are plotted in light, translucent gray, in order to display only the locality of different in-gap states in (c,d).}
	\label{fig:Hofs}
\end{figure}

\begin{figure}
	\centering
	\makebox[0pt]{\includegraphics[width=0.49\textwidth]{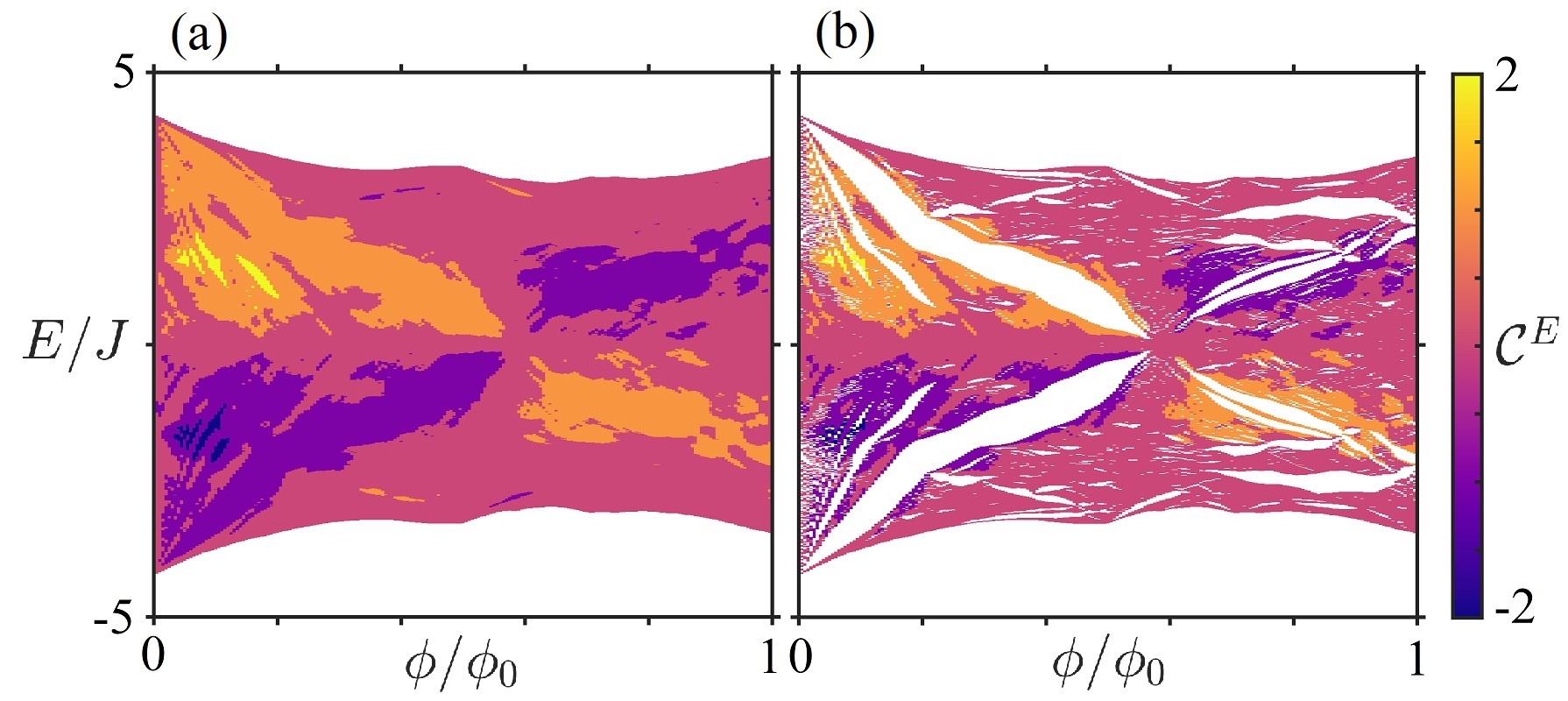}}
	\caption{Effective Chern markers for the infinite AB tiling, showing (a) the effective $\mathcal{C}^{E}$ over a range of energy values $E$ and (b) a restricted range over values in the spectrum of the infinite lattice. States in (b) with $\mathcal{C}^{E} \neq 0$ correspond to BLT states.}
	\label{fig:Hofsinf}
\end{figure}

\begin{figure}
	\centering
	\makebox[0pt]{\includegraphics[width=0.4\textwidth]{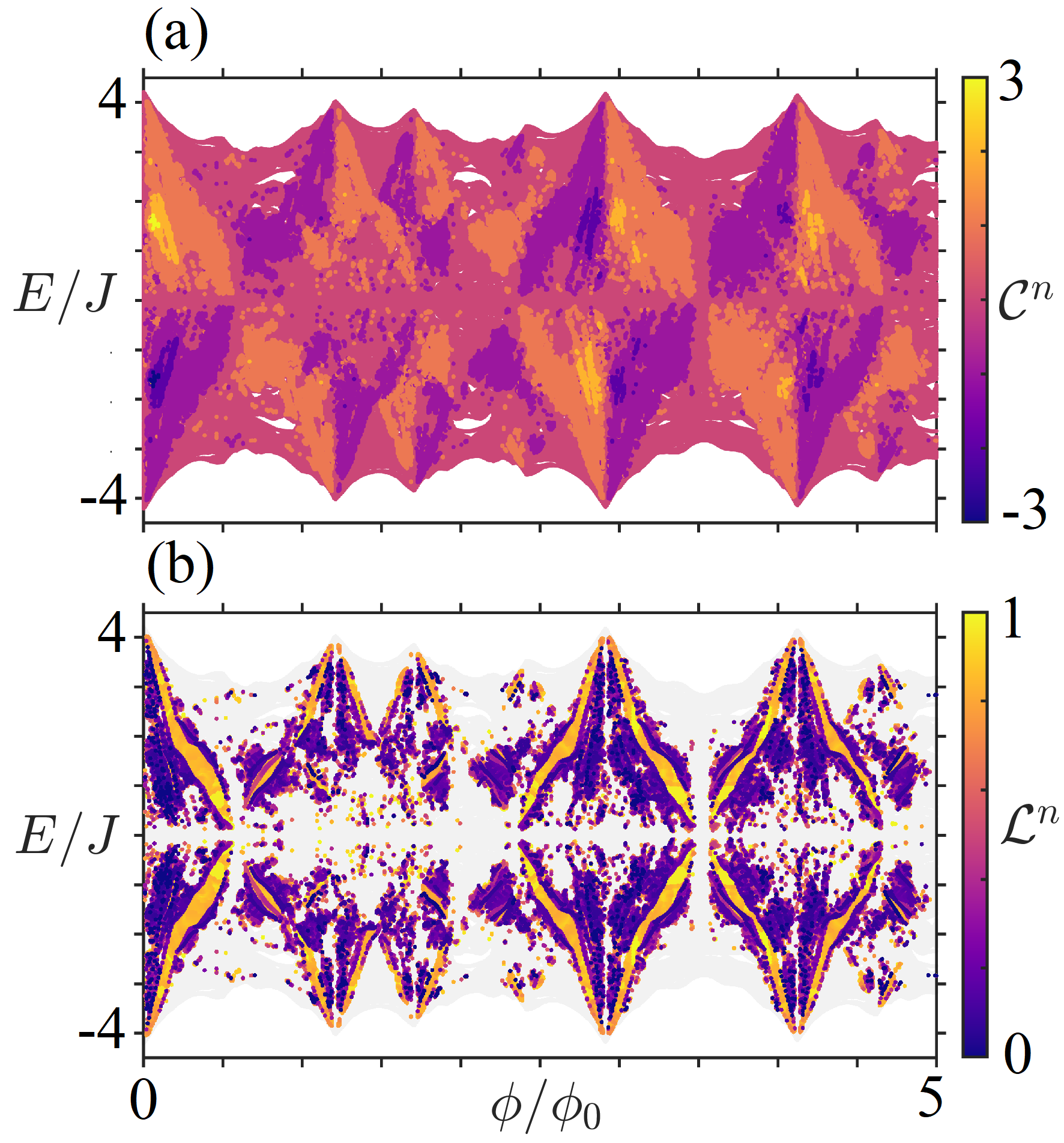}}
	\caption{Spectra of the quasicrystalline lattice, where each state is labelled according to (a) the effective Chern marker $\mathcal{C}^{n}$ and (b) the radial measure $\mathcal{L}^{n}$ across a larger range of $\phi/\phi_0$ for $N=1273$. As expected, the Hofstadter butterfly is aperiodic and has a rich, self-similar structure, with many different regions hosting BLT states.}
	\label{fig:Hofsbig}
\end{figure}

The perfect, periodic fractal nature of the Hofstadter butterfly is not preserved for quasicrystalline systems. For quasicrystals, the energy-flux plane still contains a self-similar structure, which is, in general, aperiodic. The structure of the energy-flux plane for quasicrystals has been previously studied \cite{Hatakeyama1989,Duncan2020}, and will now be utilised to show the prevalence of BLT states throughout the single-particle `phase diagram' of Hamiltonian~\eqref{eq:H}.

In Fig.~\ref{fig:Hofs}  we show the finite-size Hofstadter butterfly for two lattice sizes, with the colour map being the effective Chern marker $\mathcal{C}^n$ for each state and the corresponding radial measure of Eq.~\eqref{eq:local} for every in-gap state (i.e. non-zero $\mathcal{C}^n$). From this figure, it can be seen that there is a significant number of states that are in-gap and within the bulk, i.e. with $\mathcal{L}^n < 1$ and $\mathcal{C}^n \neq 0$. We also show the energy-flux plane for the infinite-size quasicrystal in Fig.~\ref{fig:Hofsinf}. It is clear that BLT states are present for the infinite system through large ranges of flux, with states possessing non-zero effective Chern marker being present even after ESs are removed from the system. The regions where BLT states are present in the infinite-size quasicrystal map well to those present in the finite size. 

The aperiodic Hofstadter butterflies in both the finite and infinite size show that the BLT states are not a single set of peculiar states limited to the examples shown in the previous section. Instead, the BLT states are present in the majority of parameter space for Hamiltonian~\eqref{eq:H}, with BLT states even dominating the spectrum for particular ranges of the flux. We can go to larger values of the flux, as shown in Fig.~\ref{fig:Hofsbig}, and observe even more BLT states. The prevalence of the BLT states and their variety could make the utilisation of their supported BLT particularly interesting.

\begin{figure}
	\centering
	\includegraphics[width=0.97\linewidth]{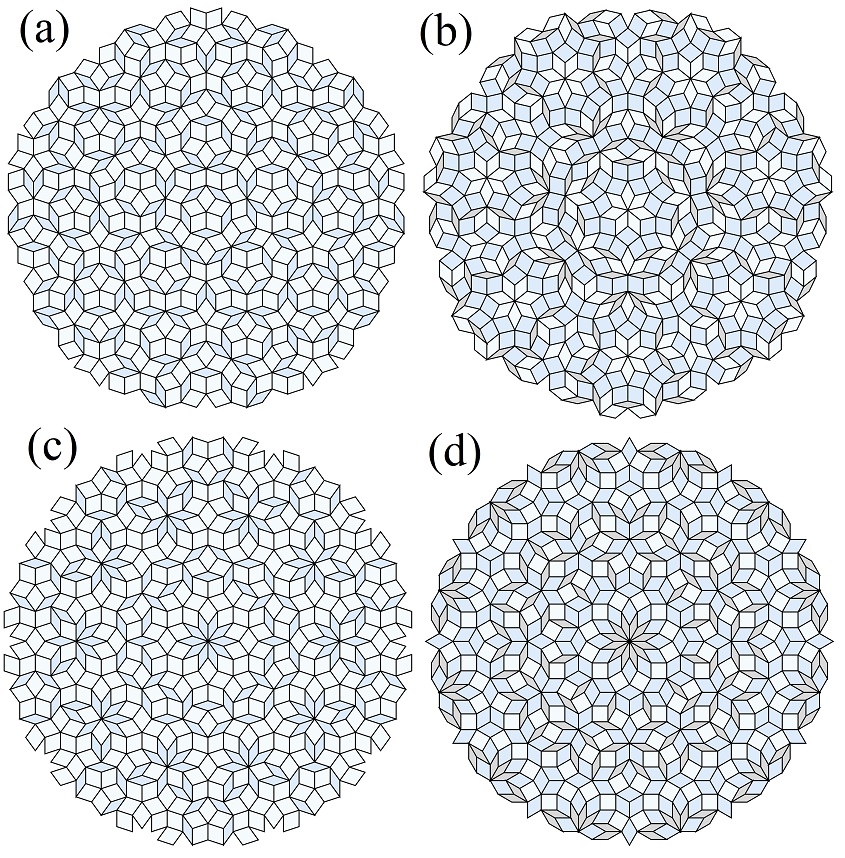}
	\caption{Finite size patches of aperiodic rhombic tilings with different rotational symmetries. Here, we consider a (a) 5-fold Moore--Penrose tiling, (b) a 7-fold tiling, (c) a 10-fold tiling and a (d) 12-fold tiling. We show $\sim 600$ sites in the corresponding vertex model for each tiling.}
	\label{fig:OtherTilings}
\end{figure}

\begin{figure}
	\centering
	\makebox[0pt]{\includegraphics[width=0.49\textwidth]{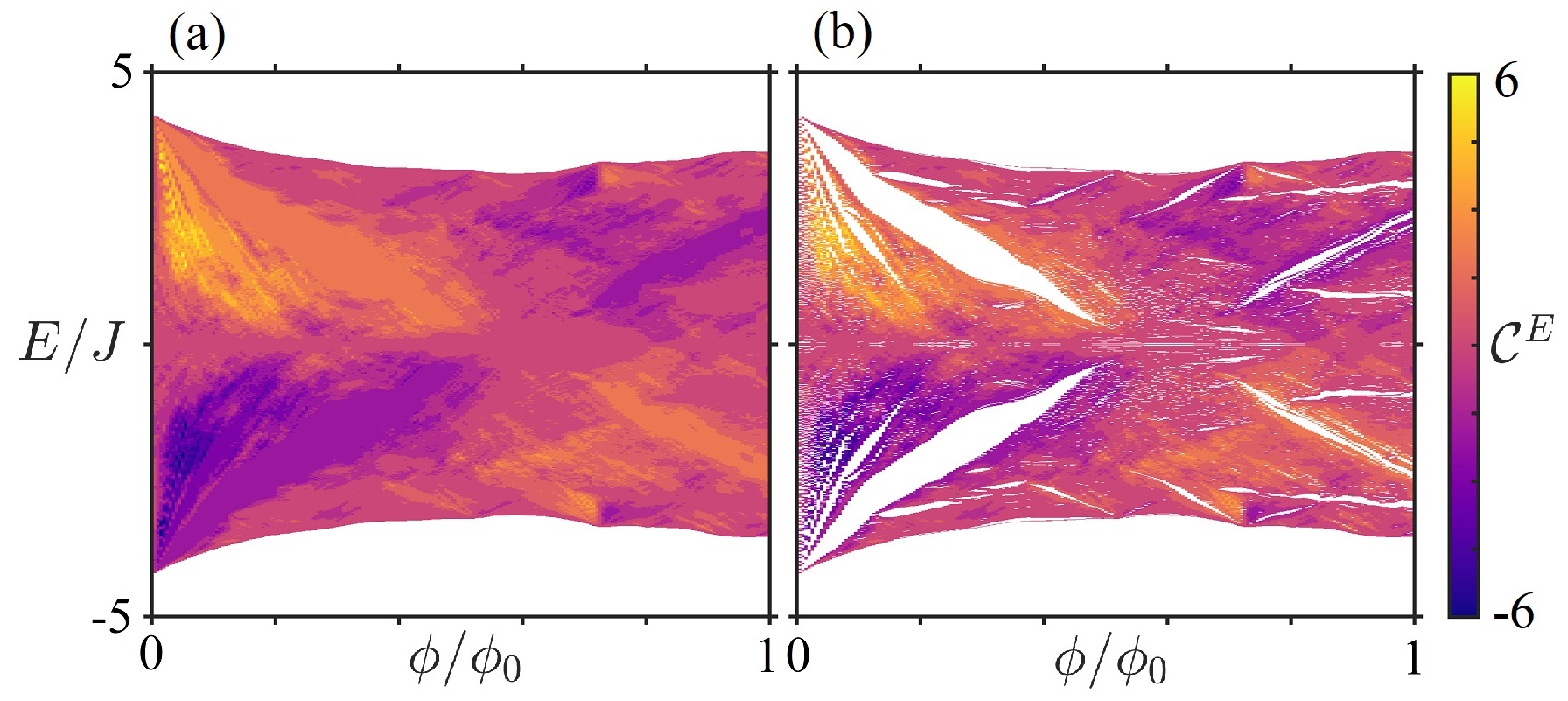}}
	\caption{Effective Chern markers for the infinite 5-Fold Moore--Penrose tiling, showing (a) the effective $\mathcal{C}^{E}$ over a range of energy values $E$ and (b) a restricted range over values in the spectrum of the infinite lattice. States in (b) with $\mathcal{C}^{E} \neq 0$ correspond to BLT states.}
	\label{fig:Hofsinf_MP}
\end{figure}

\begin{figure*}
	\centering
	\makebox[0pt]{\includegraphics[width=0.9\linewidth]{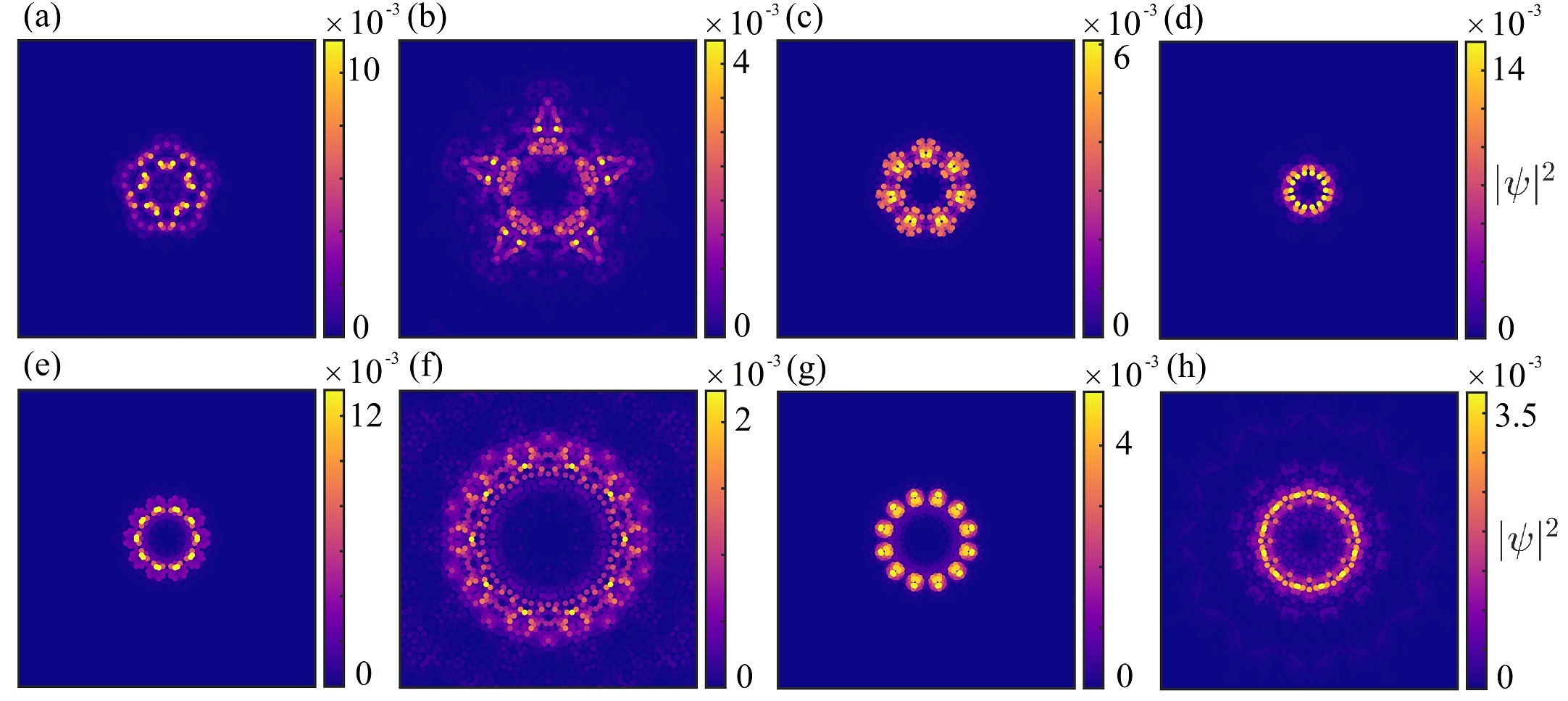}}
	\caption{Density profiles of example BLT states on infinite tilings for the (a,b) 5-fold tiling at $\phi=0.69 \phi_0$, (c,d) 7-fold tiling at $\phi=0.4 \phi_0$, (e,f) 10-fold tiling at $\phi=0.69 \phi_0$ and (g,h) 12-fold tiling at $\phi=0.4 \phi_0$. Each state corresponds to an energy value (with shown error bounds) of (a) $E=-3.0429415 \pm 10^{-7}J$, (b) $E=-1.395230\pm 2 \times 10^{-6}J$, (c) $E=-1.238795796461629\pm 5 \times 10^{-15}J$, (d) $E=-0.87206028884629\pm 10^{-14}J$, (e) $E=-0.89060992645\pm 6 \times 10^{-11}J$, (f) $E=-0.1411\pm \times 10^{-4}J$, (g) $E=-1.0537358815215\pm4 \times 10^{-13}J$ and (h) $E=-0.59818\pm 2\times 10^{-5}J$. All states shown have a non-zero effective Chern marker, and are therefore BLT states with in-gap characteristics.}
	\label{fig:BLSinfOther}
\end{figure*}

\section{Bulk Localised Transport States in Other Quasicrystals}
\label{sec:Other}

\begin{figure*}
	\centering
	\makebox[0pt]{\includegraphics[width=0.95\linewidth]{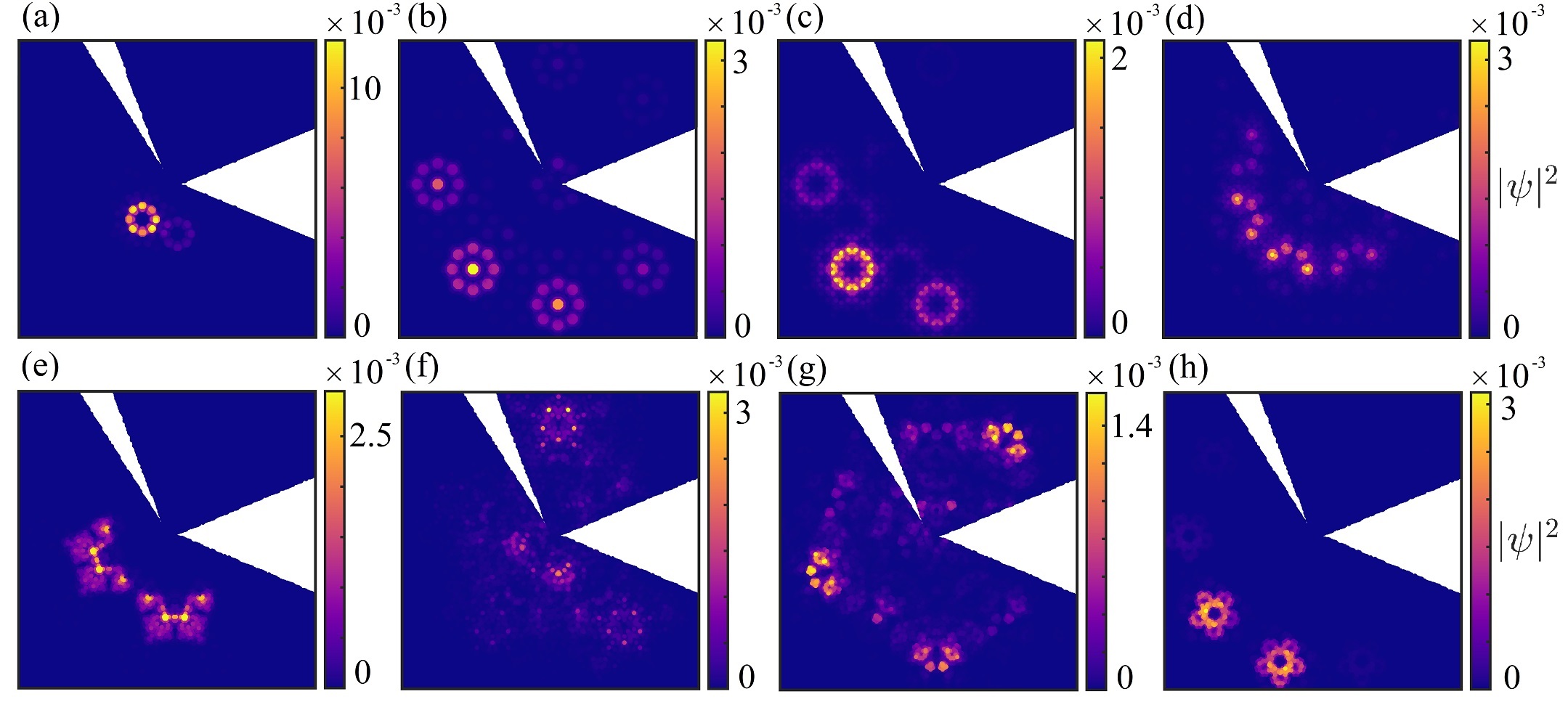}}
	\caption{Density profiles of example BLT states on infinite tilings for the symmetry broken (a-d) 8-fold tiling and (e-h) 5-fold tiling. We consider fluxes of (a,b) $\phi=0.69 \phi_0$, (c,d) $\phi=0.20 \phi_0$, (e,f) $\phi=0.69 \phi_0$ and (g,h) $\phi=0.4 \phi_0$. Each state corresponds to an energy value (with shown error bounds) of (a) $E=-1.609378\pm 3 \times 10^{-6}$, (b) $E=-0.7533\pm 2 \times 10^{-4}$, (c) $E=1.70126\pm 4 \times 10^{-5}$, (d) $E=0.9913\pm 4 \times 10^{-4}$, (e) $E=0.560996\pm 11 \times 10^{-6}$, (f) $E=2.1235\pm 4 \times 10^{-4}$, (g) $E=-1.5209\pm 3 \times 10^{-4}$ and (h) $E=-1.39457\pm 7 \times 10^{-5}$. All states shown have a non-zero effective Chern marker, and are therefore BLT states with in-gap characteristics.}
	\label{fig:BLSinfNSymm}
\end{figure*}

We have seen how prominent BLT states are within the spectra of the AB vertex model. We now show that BLT states can also populate the spectra of other kinds of quasicrystals, both with and without rotational symmetries, as long as magnetic aperiodicity, i.e. the interplay of the magnetic field with the incommensurate areas of the quasicrystal, is retained.

We plot the infinite size Hofstadter butterfly for the 5-fold lattice in Fig. \ref{fig:Hofsinf_MP}, labelled according to the effective Chern marker. This again demonstrates the removal of conventional ESs from the spectra of the 5-fold vertex model and the retention of BLT states across a broad range of flux. In other words, the appearance of BLT states is not limited to particular values of the magnetic field or the exact geometry of the lattice, and they can dominate the spectra of quasicrystals with any rotational symmetries or structure, as long as magnetic aperiodicity is retained. 

\begin{figure*}
	\centering
	\makebox[0pt]{\includegraphics[width=0.85\linewidth]{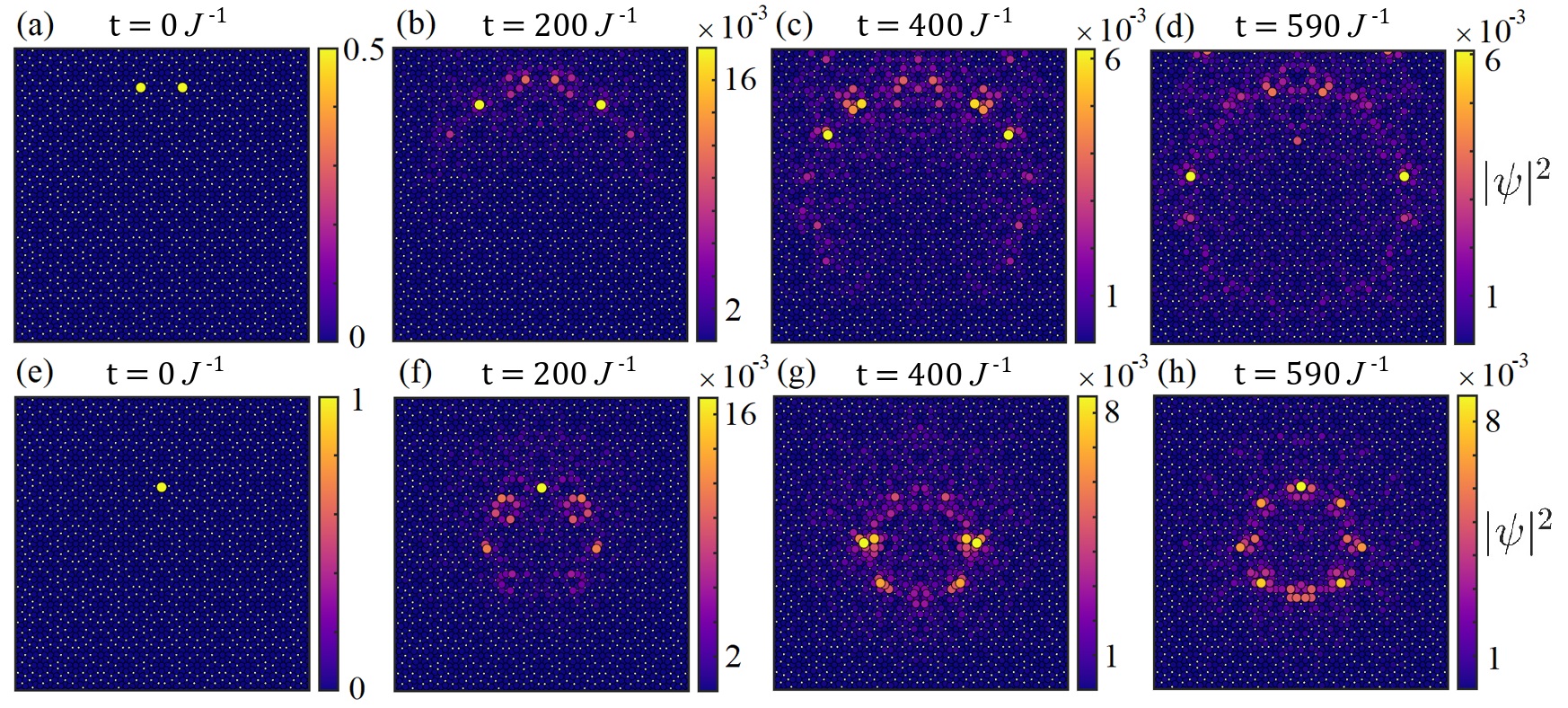}}
	\caption{Time evolution of BLT states, showing the evolution of (a-d) state $n=2987$ and (e-h) state $n=3022$ on the larger lattice of $7753$ sites. Here, each column of figures corresponds to a time frame of (a,e) $t=0J^{-1}$, (b,f) $t=200J^{-1}$, (c,g) $t=400J^{-1}$ and (d,h) $t=590J^{-1}$.  Note, that our Hamiltonian is always time independent and we do not drive the system in any way.}
	\label{fig:TransFinite}
\end{figure*}

We now show this by considering the BLT states present in vertex models of 5-fold, 7-fold, 10-fold and 12-fold rhombic tilings, which are all deduced from projections of higher-dimensional cubic lattices (see Appendix \ref{app:Tiles}). Small patches of the tilings are plotted in Fig. \ref{fig:OtherTilings} for visualisation purposes. In these examples, we also now have rhombic tilings with more than two prototiles, namely the 7-fold and 12-fold lattice. As the global rotational symmetry of the rhombic quasicrystal increases, the number of prototiles will also increase to ensure that no gaps are left in the tiling. The Hamiltonian is still that described by Eq.~\eqref{eq:H}.

In Fig. \ref{fig:BLSinfOther}, we plot several example BLT states for each Hofstadter vertex model in the infinite size. This clearly illustrates the appearance of similar-looking in-gap states to the ones observed in the 8-fold AB tiling. As expected, the structure of states on other vertex models have the rotational symmetry consistent with the tiling itself, but preservation of the global symmetry is not important to the formation of BLT states (we will show explicit examples of this later in this section). The appearance and structure of BLT states in different quasicrystals is as rich as the examples shown for the AB tiling.  All the states shown in Fig.~\ref{fig:BLSinfOther} have non-zero effective Chern marker and, hence, will support transport along them in the bulk. These results confirm that BLT states as defined in this work are a property of magnetic aperiodicity, independent of the underlying details of the lattice structure.

To demonstrate that BLT states are not a consequence of rotational (or reflection) symmetry we consider the AB and the 5-fold cases with two large arc segments removed, as shown in Fig. \ref{fig:BLSinfNSymm}. This is a rather extreme deformation to the bulk of the system, similar to the consideration of substantially deformed boundaries for regular in-gap ESs. BLT states for different flux are also shown in Fig. \ref{fig:BLSinfNSymm}, and all these states again have non-zero effective Chern marker. While the global rotational symmetry has been broken, the BLT states can still form and are prevalent throughout the spectrum, showing that they are truly a property of the bulk. The BLT states can even localise around regions with local symmetry, mimicking the BLT states observed for the non-deformed quasicrystal. Furthermore, we also observe that these BLT states may become much more highly localised to certain regions of the lattice than what is seen with global rotational symmetry.  Importantly, they can also appear in this extremely perturbed lattice as states extended and connected in both domains, as shown in Fig.~\ref{fig:BLSinfNSymm}(b,f,g), allowing for transport around the deformations, akin to the transport of regular edge states around perturbations.

\begin{figure*}
	\centering
	\makebox[0pt]{\includegraphics[width=0.8\linewidth]{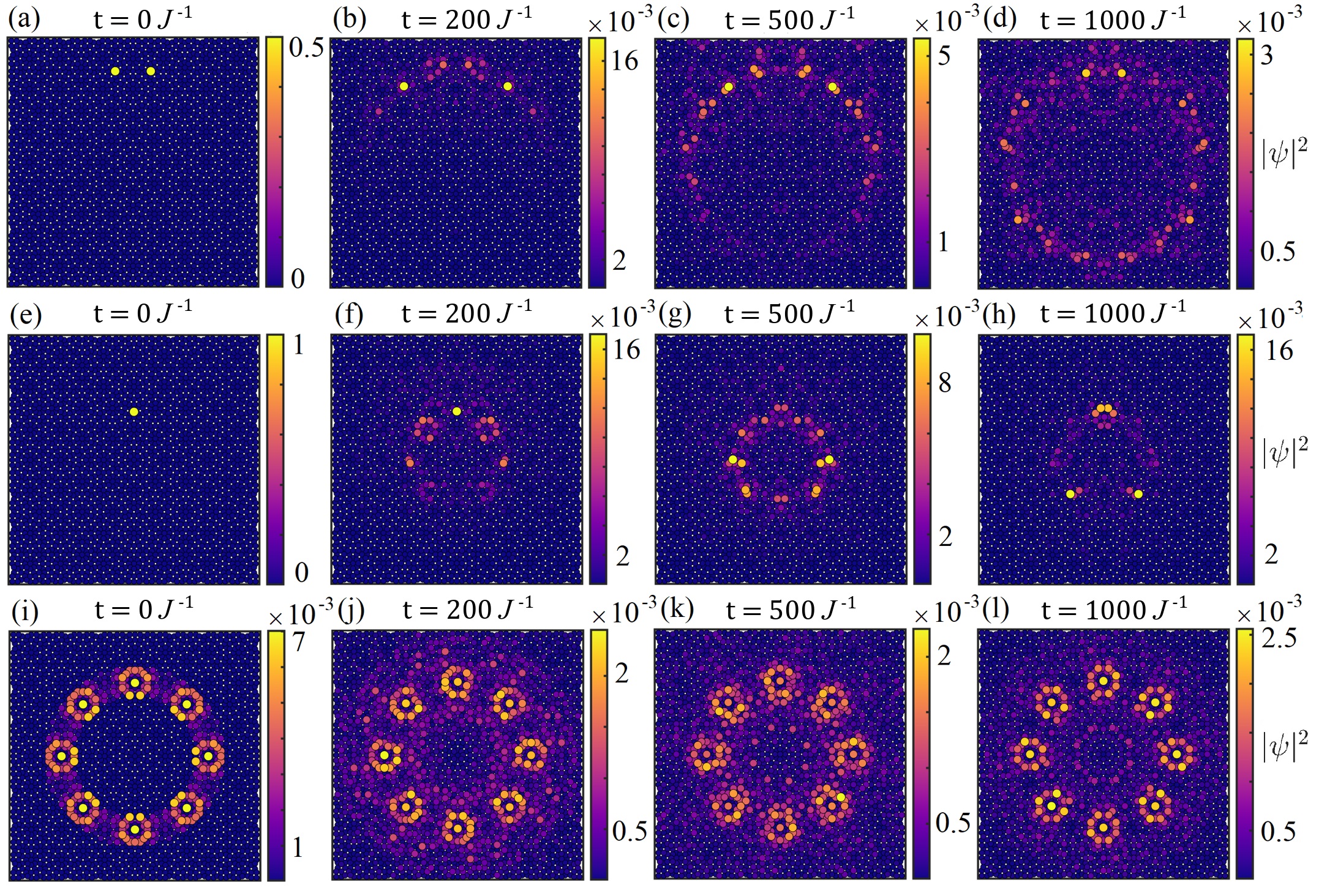}}
	\caption{Infinite-size time evolution of BLT states at time $t=0J^{-1}$ (first column), $t=200J^{-1}$ (second column), $t=500J^{-1}$ (third column) and $t=1000J^{-1}$ (fourth column). The first two rows correspond to the BLT states in Fig. \ref{fig:TransFinite} ($\phi=0.69\phi_0$), but now computed on the full infinite tiling. The third row corresponds to an excitation of the BLT state at $\phi=0.2\phi_0$ on Fig. \ref{fig:BLSinf}(c). Note, that our Hamiltonian is always time independent and we do not drive the system in any way.}
	\label{fig:TransInfinite}
\end{figure*}

\section{Bulk Localised Transport}
\label{sec:Transport}

\begin{figure*}
	\centering
	\makebox[0pt]{\includegraphics[width=0.95\linewidth]{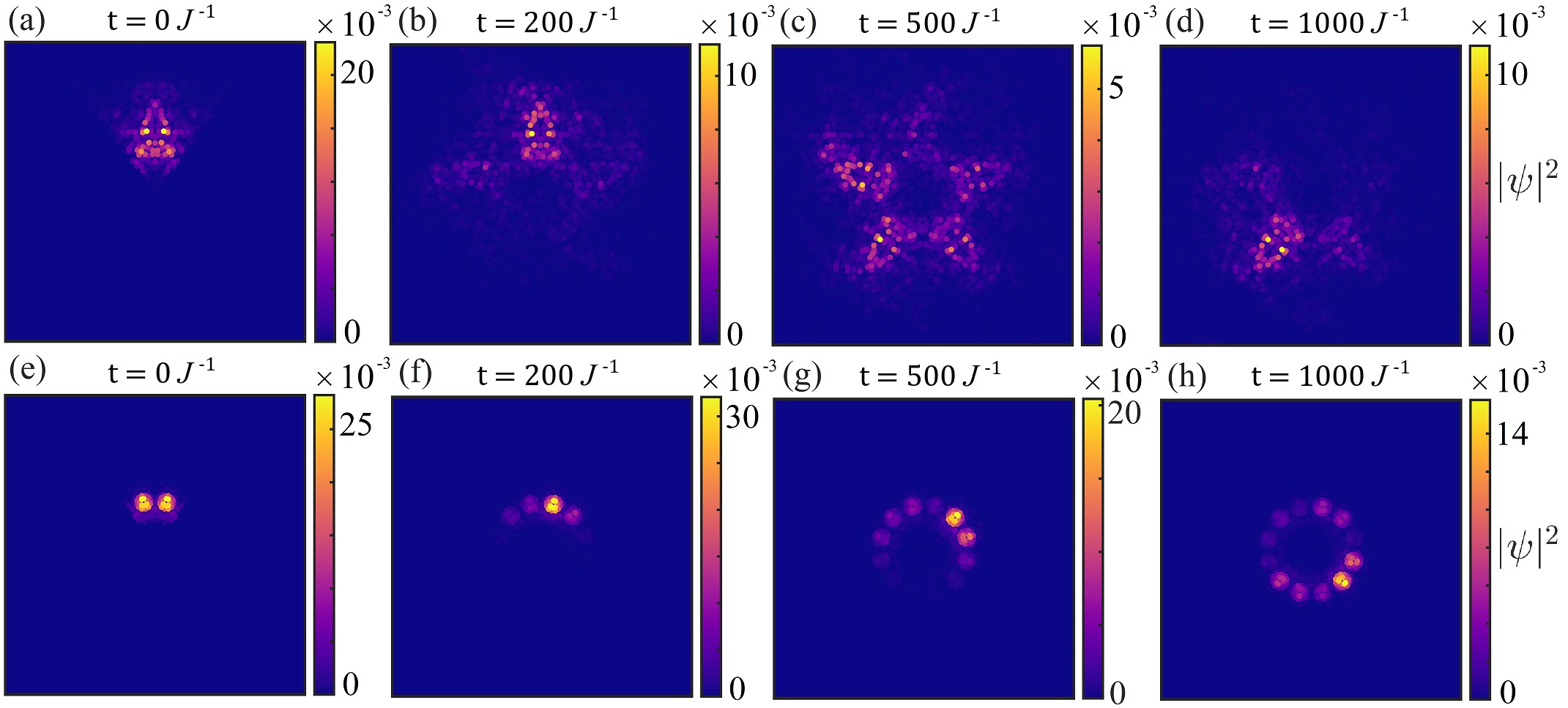}}
	\caption{Infinite-size time evolution of BLT states for other quasicrystals at time $t=0J^{-1}$ (first column), $t=200J^{-1}$ (second column), $t=500J^{-1}$ (third column) and $t=1000J^{-1}$ (fourth column). The first row corresponds to the 5-fold BLT state in Fig. \ref{fig:BLSinfOther_te}(b), with $\phi=0.69\phi_0$. The second row corresponds to an excitation of the 12-fold BLT state on Fig. \ref{fig:BLSinfOther_te}(g), with $\phi=0.4\phi_0$.}
	\label{fig:BLSinfOther_te}
\end{figure*}

For any in-gap topological state, one of the most interesting properties is their support of transport. In the case of crystalline lattices in constant magnetic fields, this is usually considered by launching a state (or particle) along the edge of the system, and observing the robust transport of a component of the state around the boundary of the system. If a boundary is instead formed within the lattice, e.g. between a topological and non-topological region, then transport can also be supported along such features due to the presence of in-gap states bound to the interface.

The BLT states found in this work are another type of in-gap state, and they should be fully expected to support transport along their locality. Indeed, we find that a quasicrystal in a magnetic field can also support long-lived BLT within the bulk, as depicted in Fig. ~\ref{fig:ExIntroFig}, due to the presence of BLT states. Several examples of BLT are shown for both the finite and infinite size in Figs.~\ref{fig:TransFinite} and~\ref{fig:TransInfinite} respectively for the AB tiling. It is clear, that a component of the initial state populates the BLT state, allowing for BLT to be supported. Note, here, we do not attempt to load into the BLT states, meaning we loose population into the other states that overlap with the initial state. For experimental scenarios, it would be prudent to instead use state preparation schemes or shortcuts to adiabaticity to load efficiently into the BLT states \cite{ostmann2017,Madail2019,Odelin2019}.

The transport shown for the infinite-size lattice in Fig.~\ref{fig:TransInfinite} is remarkable. For the infinite size, the usual transport along the edge is not present due to the lack of said edge. For the periodic infinite size, we can still have transport carrying states due to the presence of an interface or impurities. However, in the quasicrystal, an interplay of the magnetic field and quasiperiodicity can localise the particle to the bulk in the infinite size without any alterations to the lattice. These infinite size states are then as capable of carrying transport along them as any edge state in the finite system. 

We also show that this is again not a fluke of the AB Hofstadter vertex model by showing examples of BLT for some other quasicrystals in Fig.~\ref{fig:BLSinfOther_te}. These results are for the infinite size lattice and are again remarkable in their form. We show examples of BLT for the infinite tilings with 5-fold and 12-fold symmetries. Again, we observe transport supported by the BLT states which is confined within the bulk of the lattice and possible in systems regardless of the details of the boundary.

A key property of BLT could be the possibility of using its varied location throughout the lattice itself. In a crystalline finite system, the transport is only along the edge, or any internal edges imposed by the lattice structure and/or defects. However, with BLT we have shown that it is possible to have transport supported in multiple locations within the lattice for any given flux. By varying the flux, we can also tune where the BLT is supported. BLT can then give a degree of control for transport to occur within the bulk. By showing that this can be done in the infinite size, we have shown that BLT is independent of the size of the lattice, or the exact form of the boundary. This is in stark contrast to in-gap ESs, whose presence can in some cases be entirely reliant on the geometry of the edge \cite{Hatsugai2006}.

\section{Formation of Bulk Localised Transport States in a Toy Model}
\label{sec:ToyModel}

\begin{figure}
	\centering
	\makebox[0pt]{\includegraphics[width=0.45\textwidth]{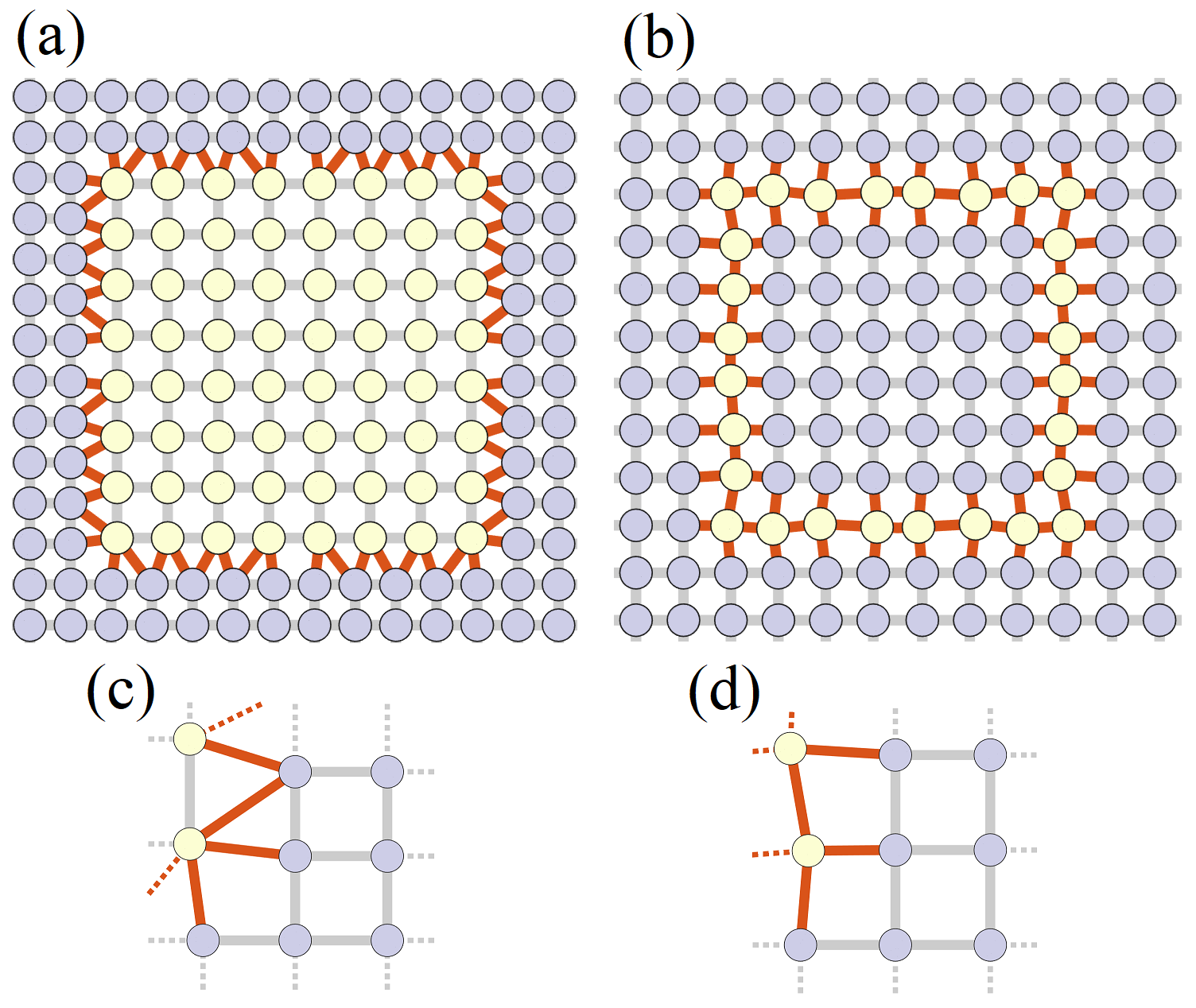}}
	\caption{Zoomed in portions of the square lattice toy models used to look at the formation of BLT states. In (a), we show a square lattice with period $l_2=\frac{13\tau}{10}$ embedded in a larger square lattice of period $l_1=1$. Connections, coloured red, are generated between the boundaries of both lattices that are less than a threshold of $1.5l_1$ apart, resulting in a coordination number dislocation. In (b), we show a square lattice with lattice spacing $l$ that has a positional disorder applied to a selection of sites. The sites coloured yellow are randomised from their original position in a small radius of $0.25l$ units. The bottom row of figures (c,d) show the bottom right corners of the same lattices in order to emphasise the incommensurate areas between different unit cells more clearly.}
	\label{fig:ModelSquare}
\end{figure}

So far, we have demonstrated that BLT states and their supported BLT are prevalent throughout the spectra and lattice of quasicrystals in magnetic fields. In this section, we will characterise how these states form through a toy model. First, we will consider a simple but misguided toy model that shows the BLT states are not formed by effective edges from the varying local coordination number. These are equivalent to dislocations being present in the system. We will then describe a simple toy model which shows that the BLT states arise due to the interplay of the quasiperiodic lattice, and mainly the irrational areas present, with the constant magnetic field.

\begin{figure}
	\centering
	\makebox[0pt]{\includegraphics[width=0.5\textwidth]{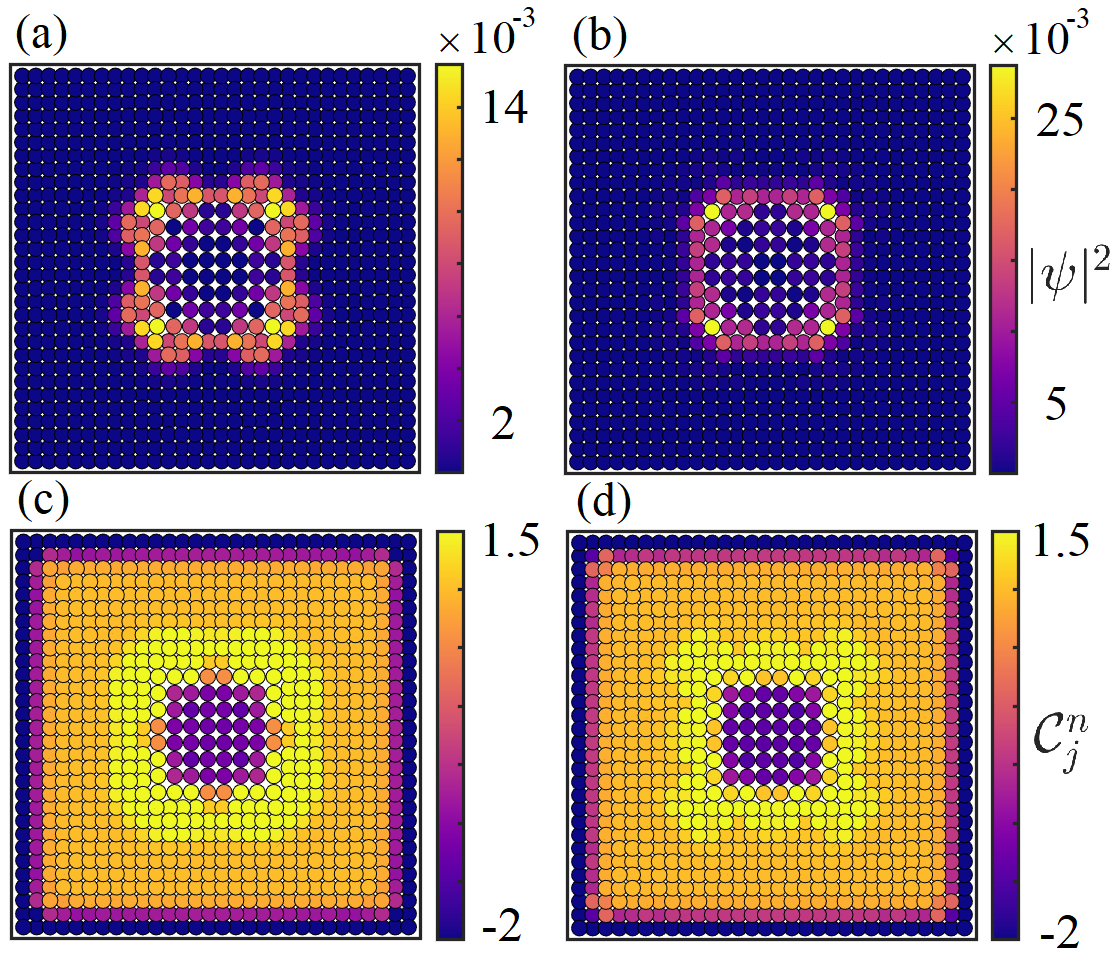}}
	\caption{Example states from the spectrum of the model in Fig. \ref{fig:ModelSquare}(a) with $\phi=0.84\phi_0$ (a-b) form across a dislocation and their (c-d) $\mathcal{C}_j^{n}$ distributions. For each case, the corresponding Bott indices are (a) $\mathcal{B}=1$ and (b) $\mathcal{B}=1$. The $\mathcal{C}_j^{n}$ distributions are saturated between $1.5$ and $-2$ for visual clarity. Each state also supports a separate bulk effective Chern marker within the inner square lattice, which is sign flipped.}
	\label{fig:Dis}
\end{figure}

\subsection{Dislocation Toy Model}

At first glance, it is easy to think that the BLT states must form only due to the quasicrystalline nature of the lattice alone. This would be through effective edges being formed in the system via the local aperiodic variation in the coordination number for each site. In many ways, this would be similar to edge states being bound to a dislocation or defect in the lattice structure. We, therefore, consider a toy model on a square lattice where we have a central region with lattice constant $l_2=\frac{13\tau}{10}$, where $\tau$ is the golden ratio, and an outer region of lattice constant $l_1=1$. The lattice constants $l_1$ and $l_2$ are incommensurate and lead to a dislocation along the boundary of these two lattices as shown in Fig.~\ref{fig:ModelSquare}(a). We couple all sites along this dislocation that are within $1.5l_1$ of each other. The dislocation is then along the interface where the coordination number varies.

Some examples of the states along the dislocation are shown in Fig.~\ref{fig:Dis}. The density profiles of these states look similar to those of the BLT states but with a crystalline 4-fold rotational symmetry, as would be expected for this toy model. They also have a corresponding non-zero Bott index and are, therefore, in-gap and will support transport. However, if we look at the local Chern marker, we observe a striking difference to the BLT states in the quasicrystal. The states along the dislocation are a product of the change of the local Chern marker across the dislocation. This in itself is not a surprise and is usually the reason why in-gap states appear on internal edges in these systems \cite{Marta2018,Pai2019,sarangi2021}. It is clear though that this is not how the BLT states form, as we do not see any change in the local Chern marker across the interface for quasicrystalline BLT states. Therefore, even though at first glance, the variation in the coordination number appears to map well to how the BLT states are formed, it cannot be their true origin.

\begin{figure}
	\centering
	\makebox[0pt]{\includegraphics[width=0.5\textwidth]{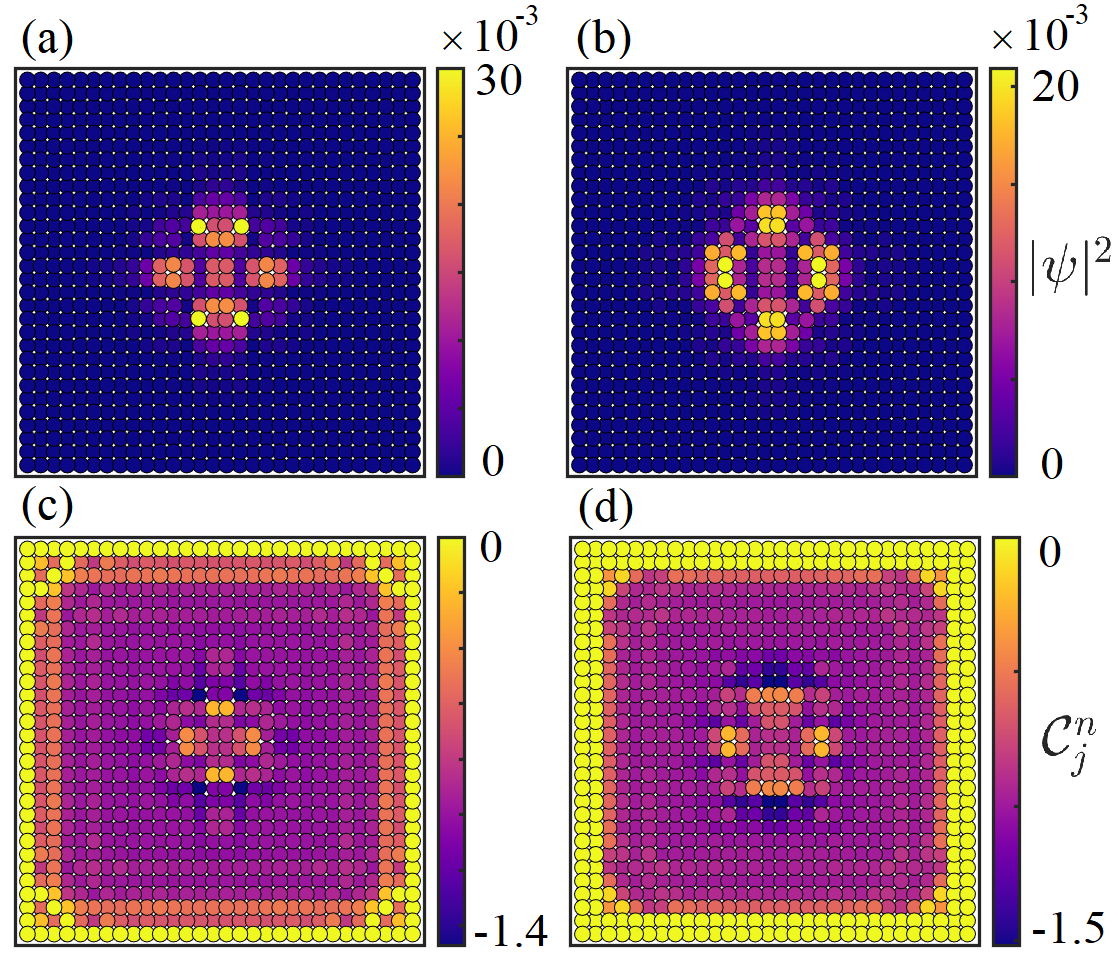}}
	\caption{Example states from the spectrum of the model in Fig. \ref{fig:ModelSquare}(b) with $\phi=0.70\phi_0$ (a-b) form around disordered fluxes and their (c-d) $\mathcal{C}_j^{n}$ distributions. For each case, the corresponding Bott indices are (a) $\mathcal{B}=-1$ and (b) $\mathcal{B}=-1$. The $\mathcal{C}_j^{n}$ distributions have the maximum value saturated to $0$ for visual clarity.}
	\label{fig:Mag}
\end{figure}

\subsection{Magnetic Aperiodicity Toy Model}
\label{sec:MagnAper}

The formation of BLT states is due to an interplay of the magnetic field with the aperiodicity of the quasicrystalline structure. This is to be expected from the results in Sec.~\ref{sec:Zoo}, where the location of the BLT states was shown to be largely dependent upon the applied magnetic field strength. Through another toy model on an originally square lattice, we can show that the magnetic aperiodicity introduced into the system is responsible for forming BLT states. In this model, we take a subset of lattice sites on a square perimeter and vary their location to a small degree, while retaining the constant coordination number and connectivity of the square lattice. An example of this toy model is shown in Fig.~\ref{fig:ModelSquare}(b), where we also enforce a 2-fold rotational symmetry for comparison. This toy model then has a small region where there is a disorder in the area of tiles. This would then map to a disorder in the flux penetrating tiles in this region, as we have cells with irrational areas to each other.

Some examples of the states along the magnetically disordered region are shown in Fig.~\ref{fig:Mag}. The states and their corresponding local Chern markers look very similar to the BLT states observed throughout this work. As expected, the states also have a 2-fold rotational symmetry and show clearly that the formation of BLT states is due to the interplay of the constant magnetic field with the self-similar aperiodic structure of the lattice. This would also explain why BLT states were not observed before in quasicrystals composed of a single tile such as the Rauzy tiling \cite{Tran2015}, as there is then no magnetic aperiodicity introduced into the system. Note, the Rauzy tiling does have a local variation in the coordination number, further supporting the conclusions from the dislocation toy model. Finally, it is worth noting that if periodic boundary conditions are applied to the hard edges of these toy models, then the BLT states remain persistent in the spectra. In other words, this again demonstrates that the formation of BLT states is not linked to the presence or geometry of the system's boundary.

To observe BLT states, it is therefore key to have the aperiodic nature of the lattice compete with the magnetic field. The simplest way to do this is to have a lattice constructed from building blocks that have incommensurate areas to one another, like the two prototiles of the AB tiling. We would then expect that BLT states are present in all quasicrystalline lattices which do not have a single, or multiple commensurate building blocks. We would also expect that the presence of BLT states should be independent of the vector potential, as long as the choice does not destroy the interplay with the incommensurate building blocks of the lattice.

\section{Conclusions}

We have shown that the conventional picture of insulators, metals, and topological insulators with surface states is not the full story for quasiperiodic systems. When the quasiperiodic nature of the quasicrystalline lattice interacts with the magnetic field, it is possible for states to form that are unique in their character. These states are localised to the bulk, but are in-gap and support transport along them in the bulk. The BLT states are far from a peculiarity and can exist throughout the spectra of 2D quasicrystals in magnetic fields due to a magnetic aperiodicity. Magnetic aperiodicity is not an artificial construct and is natural in the majority of 2D quasicrystals, due to the incommensurate nature of the building blocks of the lattice. For the case of the vertex models studied in this work, this incommensurate nature arises from the incommensurate areas of the prototiles.

BLT states are even sometimes the dominant in-gap state for large finite system sizes. We have confirmed that the BLT states can exist in a variety of regions within the bulk of the lattice, with their position being dependent on the flux, or equivalently the magnetic field strength. Through the use of a new numerical technique, we have also shown that BLT states are present in the infinite-size lattice. This is quite remarkable as they are in-gap states, with corresponding non-zero topological measures. By exciting regions within the bulk of the lattice, we have also shown that BLT is supported in both the finite and infinite size.

A future direction of interest could be to realise BLT states of quasicrystals in an experimental setting. While there are current developments of realising quasicrystalline problems in cold atoms \cite{Gopalakrishnan2013,Viebahn2019}, a promising setting for current realisation of the BLT states could be photonic lattices \cite{GARANOVICH20121}. Photonic lattices allow a high degree of controllability in the lattice geometry and are favourable for realising the BLT states due to a large number of sites being possible. The lattice is usually etched into the two-dimensional plane of a fused silica crystal, with time then being transposed onto the third dimension of the crystal. Photonic lattices are good simulators for single-particle physics, and synthetic gauge fields can be realised via helical waveguides \cite{lumer2019}, a Floquet step-like approach \cite{mukherjee2017,Goldman2015}, or a strain across the lattice \cite{Guglielmon2021}. Furthermore, there has also been a recent proposal to realise a Penrose quasicrystal in a synthetic vector potential using helical waveguides \cite{Bandres2016}. As the key to the realisation of BLT states is the magnetic aperiodicity, we would expect that any of these current techniques could potentially be utilised to probe the physics of BLT states.

The discovery of a zoo of BLT states in quasicrystals opens a number of intriguing open problems and future research directions. One possible line of work is to consider the nature of the spectra for the infinite-size tiling. Usually, we would consider these states to all be ordinary bulk states, but as we have shown this is not necessarily the case, even without any impurities or true disorder in the system. Another interesting question is how the BLT states would appear or alter the physics in the presence of interactions. The study of the physics of two-dimensional quasicrystals including interactions is an emergent topic \cite{Johnstone2019,johnstone2020,Gautier2021}. With the recent advances in many-body numerical techniques \cite{SCHOLLWOCK2011,ORUS2014,Silvi2019}, the physics of BLT states and their ramifications in quasicrystalline lattices in the many-body regime could soon be probed.

The reliance on magnetic aperiodicity also initiates an interesting set of questions. This is not a sole property of quasicrystals and could be incorporated into crystal structures like that considered briefly to generate an aperiodic Hofstadter butterfly in Ref.~\cite{Duncan2020}, or extensions of the toy model of magnetic aperiodicity considered in this work. This approach could be used to design effective regions in the system where BLT states are desired to localise, through the appearance of incommensurate phases.  These regions could then be tuned to support or not support transport through changes in the applied magnetic field strength. This would potentially allow the design and control of specific regions supporting localised transport within the lattice bulk, with interesting applications.

The understanding of the physics of electronic-like states in quasicrystals is at an early stage of its development, especially concerning potential applications to quantum technologies. There is much work to be done to realise the potential of these fascinating and complex structures. This work has shown one possible exotic behaviour of these systems in magnetic fields; the existence of BLT states and their supported transport. Future work to understand the applications of BLT states to quantum problems, like the applications considered for typical edge states \cite{Thouless1982,Hasan2010,Xiao2010,lang2017}, could be particularly fruitful.

\begin{acknowledgments}
The authors acknowledge helpful discussions with Andrew J. Daley, Terry A. Loring, Manuel Valiente, and Alexander Watson. D.J. acknowledges support from EPSRC CM-CDT Grant No. EP/L015110/1. M.J.C. acknowledges support from a Research Fellowship at Trinity College, Cambridge. A.E.B.N. and C.W.D. acknowledges support from the Independent Research Fund Denmark under Grant Number 8049-00074B. C.W.D. acknowledges support by the EPSRC Programme Grant DesOEQ (EP/P009565/1),  by the European Union's Horizon 2020 research and innovation program under grant agreement No.~817482 PASQuanS, and the EPSRC Quantum Technologies Hub for Quantum Computing and Simulation (EP/T001062/1).
\end{acknowledgments}

\appendix

\section{Details of the Infinite-Size Algorithms}
\label{app:Inf}

Here we provide details of the algorithms that are used to tackle infinite-dimensional spectral problems. We split the discussion into three subsections corresponding to each computed spectral quantity.

\subsection{Computing Spectra with Error Control}

We will describe the algorithm for infinite, sparse (finitely many non-zero entries in each column) matrices representing Hermitian Hamiltonians. For non-sparse matrices and even non-Hermitian operators, see Ref.~\cite{Colbrook2019}. Extensions to unbounded operators and partial differential operators can be found in Ref.~\cite{colbrook2019b}. 

Recall that in our setting, the Hamiltonian $H$ can be represented by an infinite Hermitian matrix, $\hat H=\{\hat H_{ij}\}_{i,j\in\mathbb{N}}$ and we are given a function $f:\mathbb{N}\rightarrow\mathbb{N}$ such that $\hat H_{ij}=0$ if $i>f(j)$, thus describing the sparsity of $\hat H$. For $z\in\mathbb{R}$, the key quantity to compute is
\begin{equation}
F_n(z):=\sigma_1(P_{f(n)}(\hat H-z)P_n),
\end{equation}
where $P_m$ denotes the orthogonal projection onto the linear span of the first $m$ basis vectors and $\sigma_1$ denotes the smallest singular value of the corresponding rectangular matrix. The function $F$ is an upper bound for the distance of $z$ to the spectrum $\mathrm{Sp}(H)$, and converges down to this distance uniformly on compact sets as $n\rightarrow\infty$. There are numerous ways to compute $F_n$, such as standard iterative algorithms or incomplete Cholesky decomposition of the shifts $P_n(\hat H-z)P_{f(n)}(\hat H-z)P_n$ (see the supplementary material of \cite{Colbrook2019} for a discussion). The other ingredient is a grid of points $G_n=\{z_1^{(n)},...,z_{j(n)}^{(n)}\}\subset\mathbb{R}$ providing the wanted resolution $r_n$ over the spectral region of interest.

The algorithm is sketched in Algorithm \ref{alg:spec_comp}, where $\tilde F_n$ denotes the described suitable approximation of $F_n$ (which can be computed in parallel). The simple idea of the method is a local search routine. If $\tilde F_n(z)\leq 1/2$, we search within a radius $\tilde F_n(z)$ around $z$ to minimise the approximated distance to the spectrum. This gives the set $M_z$ which is our best estimate of points in the spectrum near $z$. The output is then the collection of these local minimisers. The algorithm's output, $\Gamma_n(H)$, converges to the spectrum $\mathrm{Sp}(H)$ of the full infinite-dimensional operator as $n\rightarrow\infty$ (for suitable $r_n\rightarrow\infty$). Note that this convergence is free from edge states. Moreover, the error bound of the algorithm satisfies
\begin{equation}
\sup_{z\in\Gamma_n(H)} \mathrm{dist}(z,\mathrm{Sp}(H))\leq E_n
\end{equation}
and the output $E_n$ converges to zero as $n\rightarrow\infty$ (proven in Ref.~\cite{Colbrook2019}). For a desired accuracy $\delta>0$, we simply increase $n$ until $E_n\leq\delta$. In our numerical experiments we chose $\delta$ to be smaller than the required spectral resolution. Finally, the output $V_n$ consists of the approximate states corresponding to the output $\Gamma_n$.

\begin{algorithm}[H]
\vspace{2mm}
\textbf{Input:} $\hat H$, $f$, $n$ and $G_n$ (with resolution $r_n$).\\
\begin{algorithmic}[1]
\STATE For $z\in G_n$, approximate $F_n(z)$ to accuracy $(2r_n)^{-1}$ from above. Call the approximation $\tilde F_n(z)$ and assume it takes values in $(2r_n)^{-1}\mathbb{Z}$.
\STATE For $z\in G_n$, let $v_n(z)$ denote the approximation of the right-singular vector of $P_{f(n)}(\hat H-z)P_n$ corresponding to the smallest singular value.
\STATE For $z\in G_n$, if $\tilde F_n(z)\leq 1/2$, then set
\begin{align*}
I_z&=\{w\in G_n:|w-z|\leq \tilde F_n(z)\},\\
M_z&=\left\{w\in G_n:\tilde F_n(w)=\min_{x\in I_z}\tilde F_n(x)\right\}.
\end{align*}
Otherwise, set $M_z=\emptyset$.
\end{algorithmic} \vspace{2mm} \textbf{Output:} $\Gamma_n=\cup_{z\in G_n}M_z$ (approximation of spectrum), $E_n=\max_{z\in\Gamma_n}\tilde F_n(z)$ (error bound) and $V_n=\cup_{z\in\Gamma_n}\{v_n(z)\}$ (approximate states).
\caption{
Computation of spectrum and the associated approximate states with error control. The computation of $\tilde F_n$ can be performed in parallel.
}\label{alg:spec_comp}
\end{algorithm}

\subsection{Computing Spectral Measures and Local Chern Markers}\label{alg_desc_chern}

In this section, we will assume access to a routine that approximates the action of the resolvent $(H-z)^{-1}$ on a vector with error bounds. In the scenario of the current paper, this can be done through the rectangular truncations $P_{f(n)}(\hat H-z)P_n$ and solving the resulting overdetermined linear system in the least squares sense. The residual converges to zero as $n\rightarrow\infty$ and can be used to provide the needed error bounds through an adaptive selection of $n$ (see~\cite[Th. 2.1]{colbrook2019computing}).

We use the high-order kernel machinery developed in Ref.~\cite{colbrook2020}, where the following definition is made.
\begin{definition}[$m$th order kernel]
\label{def:mth_order_kernel}
Let $m\in\mathbb{N}$ and $K\in L^1(\mathbb{R})$. We say $K$ is an $m$th order kernel if:
\begin{itemize}
	\item[(i)] Normalised: $\int_{\mathbb{R}}K(x)dx=1$.
	\item[(ii)]  Zero moments: $K(x)x^j$ is integrable and $\int_{\mathbb{R}}K(x)x^jdx=0$ for $0<j<m$.
	\item[(iii)] Decay at $\pm\infty$: There is a constant $C_K$, such that $
	\left|K(x)\right|\leq {C_K}{(1+\left|x\right|)^{-(m+1)}}$, $\forall x\in \mathbb{R}.$
\end{itemize}
\end{definition}
We set $K_{\epsilon}(\cdot)=\epsilon^{-1}K(\cdot /\epsilon)$. In Ref.~\cite{colbrook2020}, theorems were proven on the rates of approximating a probability measure $\mu$ by the convolution $K_{\epsilon}*\mu$. For $m$th order kernels, under suitable conditions, $m$th order convergence in $\epsilon$ holds (i.e. the error scales as $\epsilon^m$ up to logarithmic factors). These rates can be carried over to approximating the projection-valued measure $\mathcal{E}$ described in Sec. \ref{sec:Top}.

High-order kernels can be constructed using rational functions as follows. Let $\{a_j\}_{j=1}^m$ be distinct points in the upper half plane and suppose that the constants $\{\alpha_j\}_{j=1}^m$ satisfy the following (transposed) Vandermonde system:
\begin{equation}\label{eqn:vandermonde_condition}
\begin{pmatrix}
1 & \dots & 1 \\
a_1 & \dots & a_{m} \\
\vdots & \ddots & \vdots \\
a_1^{m-1} &  \dots & a_{m}^{m-1}
\end{pmatrix}
\begin{pmatrix}
\alpha_1 \\ \alpha_2\\ \vdots \\ \alpha_{n_1}
\end{pmatrix}
=\begin{pmatrix}
1 \\ 0 \\ \vdots \\0
\end{pmatrix}.
\end{equation}
Then the kernel
\begin{equation}
K(x)=\frac{1}{2\pi i}\sum_{j=1}^{n_1}\frac{\alpha_j}{x-a_j}-\frac{1}{2\pi i}\sum_{j=1}^{n_2}\frac{\overline{\alpha_j}}{x-\overline{a_j}},
\end{equation}
is an $m$th order kernel \cite{colbrook2020}, and we have the following generalisation of Stone's formula
\begin{equation}
\label{gen_stone_comp}
[K_{\epsilon}*\mathcal{E}](x)=\frac{-1}{2\pi i}\sum_{j=1}^{m}\left[\alpha_j (H-(x-\epsilon a_j))^{-1}-c.c.\right].
\end{equation}
As a natural extension of the Poisson kernel, whose two poles are at $\pm i$, we consider the choice
\begin{equation}\label{eqn:equi_poles}
a_j=\frac{2j}{m+1}-1+i, \qquad 1\leq j\leq m.
\end{equation}
We then determine the residues by solving the Vandermonde system in Eq.~\eqref{eqn:vandermonde_condition}. The first six kernels are explicitly written down in Table \ref{tab_kernel}.

\begin{table*}
\renewcommand{\arraystretch}{1.6}
\centering
\begin{tabular}{l|c|c}
$m$ & $\pi K(x)\prod_{j=1}^m(x-a_j)(x-\overline{a_j})$ & $\{\alpha_1,\ldots,\alpha_{\ceil{m/2}}\}$\\
\hline
$2$ & $\frac{20}{9}$ & $\left\{\frac{1+3i}{2}\right\}$\\
$3$ &$-\frac{5}{4}x^2+\frac{65}{16}$ & $\left\{-2+i,5\right\}$\\
$4$ & $-\frac{3536}{625}x^2+\frac{21216}{3125}$ & $\left\{\frac{-39-65i}{24},\frac{17+85i}{8}\right\}$\\
$5$ & $\frac{130}{81}x^4 - \frac{12350}{729}x^2 + \frac{70720}{6561}$ & $\left\{\frac{15-10i}{4},\frac{-39+13i}{2},\frac{65}{2}\right\}$\\
$6$ & $\frac{1287600}{117649}x^4 - \frac{34336000}{823543}x^2 + \frac{667835200}{40353607}$ & $\left\{\frac{725+1015i}{192},\frac{-2775-6475i}{192},\frac{1073+7511i}{96}\right\}$\\
\end{tabular}
\caption{The numerators and residues of the first six rational kernels with equispaced poles (see Eq.~\eqref{eqn:equi_poles}). We give the first ${\ceil{m/2}}$ residues because the others follow by the symmetry $\alpha_{m+1-j}=\overline{\alpha_j}$.\label{tab_kernel}}\renewcommand{\arraystretch}{1}
\end{table*}

With this in hand, and for a given energy value $E$, we compute the smoothed spectral projections in Eq.~\eqref{eq:Proj2} using the trapezoidal rule. The quantities $\hat{x}_{\epsilon}^{E}$ and $\hat{y}_{\epsilon}^{E}$ can be computed via successive applications of the relevant projectors. This is outlined in Algorithm \ref{alg:chern_comp}, which computes the local Chern markers over a grid of energy values of spacing $\Delta E$. In practice, the algorithm has two levels of parallelism. We can compute resolvents in parallel across different energy values $E_j$ and we can perform the algorithm in parallel for different sites indexed by $i$.

\begin{algorithm}[H]
\vspace{2mm}
\textbf{Input:} $\hat H$, $f$, $m\in\mathbb{N}$ (order of kernel), $\epsilon>0$ (smoothing parameter), $\Delta E$ (energy or spectral spacing), $i$ (site index), $A_c$ (reference area of the lattice), $L$ (lower bound for $\mathrm{Sp}(H)$) and $M\in\mathbb{N}$ (number of energy values).\\
\begin{algorithmic}[1]
\STATE Set $E_j=L+j\times\Delta E$ for $j=0,1,...,M$ and for $j=1,...,M$, set
$$
\hat P_{\epsilon,\Delta E}^{E_j}=\Delta E\sum_{k=0}^j\frac{[K_{\epsilon}*\mathcal{E}](E_{k-1})+[K_{\epsilon}*\mathcal{E}](E_k)}{2},
$$
where $K$ is the $m$th order kernel from Table \ref{tab_kernel}, and the resolvents $(H-z)^{-1}$ in Eq.~\eqref{gen_stone_comp} are computed adaptively through rectangular truncations corresponding to the function $f$ (see main text).
\STATE Set $\hat Q_{\epsilon,\Delta E}^{E_j}=\mathbb{I}-\hat P_{\epsilon,\Delta E}^{E_j}$, where $\mathbb{I}$ denotes the identity operator.
\STATE Define the operators
$$
\hat{x}_{\epsilon,\Delta E}^{E_j} = \hat{Q}^{E_j}_{\epsilon,\Delta E} \hat{x} \hat{P}^{E_j}_{\epsilon,\Delta E}, \, \, \, \hat{y}_{\epsilon,\Delta E}^{E_j} = \hat{P}^{E_j}_{\epsilon,\Delta E} \hat{y} \hat{Q}^{E}_{\epsilon,\Delta E}.
$$
\end{algorithmic}  \vspace{2mm}\textbf{Output:} Local Chern markers
$$
\mathcal{C}_i^{E_j} = \dfrac{-4\pi}{A_c^2} \Imm \left\{ \bra{i} \hat{x}_{\epsilon,\Delta E}^{E_j} \hat{y}_{\epsilon,\Delta E}^{E_j} \ket{i}\right\}
$$
at energy value $E_j$ for $j=1,...,M$.%\vspace{-2mm}
\caption{
Computation of local Chern markers. There are two levels of parallelism: the computation of resolvents and the entire algorithm for different sites indexed by $i$.
}\label{alg:chern_comp}
%\end{flushleft}
\end{algorithm}

\subsection{Computing Transport Properties}

Finally, we will discuss the computation of transport properties. Given an initial wavefunction $\psi_0$, we wish to compute
\begin{equation}\label{func_calc_desc}
\begin{split}
\psi(t)&=\exp(-i Ht)\psi_0\\
&=\frac{1}{2\pi i}\int_{\gamma}\exp(-i zt)\left[(H-z)^{-1}\psi_0\right]dz,
\end{split}
\end{equation}
where $\gamma$ is a closed contour looping once around the spectrum. Suppose that the spectrum is located in an interval $[a,b]\subset\mathbb{R}$. We take $\gamma$ to be a rectangular contour split into four line segments: two parallel to the imaginary axis with real parts $a-1$ and $b+1$ and two parallel to the real axis with imaginary parts $\pm \eta$ ($\eta>0$). Along these line segments we apply Gaussian quadrature with enough quadrature nodes for the desired accuracy (the number of nodes can be found by bounding the analytic integrand). Suppose that the weights and nodes for the quadrature rule applied to the whole of $\gamma$ are $\{w_j\}_{j=1}^N$ and $\{z_j\}_{j=1}^N$. Then the approximation of (\ref{func_calc_desc}) is given by
\begin{equation}
\psi(t)\approx \sum_{j=1}^N \frac{w_j}{2\pi i} \exp(-i z_jt)\left[(H-z_j)^{-1}\psi_0\right].
\end{equation}
The $(H-z_j)^{-1}\psi_0$ are computed using the adaptive method outlined in Sec. \ref{alg_desc_chern}, which can be performed in parallel across the different quadrature nodes. We also reuse these computed vectors for different times $t$ (numerically, this requires $\eta$ to not be too large and suitable $N$ can be selected for a finite interval of desired times $t$).

\section{Construction of Aperiodic Tilings}
\label{app:Tiles}
The quasicrystalline tilings considered in this work are deduced from 2D cut-and-project sets of a $D$ dimensional lattice $\mathcal{Z}$. With this method, the key idea is to map lattice points in $\mathbb{R}^D \rightarrow \mathbb{R}^2$. We consider rhombic tilings in this work, which requires that the $\mathcal{Z}$ are hypercubic structures. The set of basis vectors of $\mathcal{Z}$ are then simply the permutations of $1$ in $\mathbb{R}^D$ zero vectors. In order to produce the tiling, we first define a rotation on the points $\vec{V} \in \mathcal{Z}$ relative to the origin $\vec{0}$
\begin{equation} \label{eq_RMV}
\begin{aligned}
\vec{W} = \mathcal{R}\vec{V},
\end{aligned}
\end{equation}
where $\vec{V}$ is defined by the basis vectors of $\mathcal{Z}$, $\mathcal{R}$ is an incommensurate rotation operator and $\vec{W}$ is a transformed position. For hypercubic lattices, the columns of $\mathcal{R}$ naturally define the transformed basis vectors, leading to the following constraints on $\mathcal{R}$
\begin{equation}
\begin{aligned}
\textbf{C}_i \cdot \textbf{C}_j = \delta_{ij},
\end{aligned}
\end{equation}
where $\textbf{C}_i$ is the $i$th column of $\mathcal{R}$ in vector form and $|\mathcal{R}|=1$ for a rotation operator. This is effectively a statement that the normalisation and orthogonality of basis vectors should be invariant under rotation. We can now define two unique subspaces of the $\mathbb{R}^D$ superspace of $\mathcal{Z}$. First, we have the tiling space $\mathcal{T} \in \mathbb{R}^2$, which is the $2$ dimensional projection of the transformed points $\vec{W}$. In other words, the points in $\mathcal{T}$ are defined by the first two elements of $\vec{W}$. Second, we have the internal space $\mathcal{I} \in \mathbb{R}^{D-2}$, which is the $D-2$ dimensional projection of $\vec{W}$. Similar to before, a point in $\mathcal{I}$ is defined by the next $D-2$ elements of $\vec{W}$, after the points $\mathcal{T}$.

To now form the tiling, we also define a cutoff for points in $\mathcal{I}$. This is done by taking the projected unit cell of $\mathcal{Z}$ as a bounding volume for points in $\mathcal{I}$. The reason this is necessary is due to the fact that the projection of all $\vec{W}$ to $\mathcal{T}$ will lead to a dense set of points, with no quasicrystalline structure and many overlapping tiles. Therefore, the final tiling in $\mathcal{T}$ now only accepts points whose dual in $\mathcal{I}$ is bounded by the projected unit cell, ensuring no tiles overlap.

For each tiling generated from a $D$ dimensional hypercube, there may also exist a family of isomorphism classes. These are tilings with equivalent or up to a $2 \times D$ rotational symmetry, but with different local properties and frequency of tiles. These can be found by simply offsetting projected points in $\mathcal{I}$ by a small vector $\vec{s}$. 

Finally, we note here that the tilings considered in this work are generated from hypercubes in $D = 4$ (8-fold), $D = 5$ (5-fold and 10-fold), $D = 6$ (12-fold) and $D = 7$ (7-fold).

\vspace{1cm}

\bibliographystyle{apsrev4-2}
\bibliography{EdgeInternal}

\end{document}